\documentclass[conference]{IEEEtran}
 \usepackage{multirow} 
\usepackage{cite}
\usepackage{comment}
\usepackage{amsmath,amssymb,amsfonts}
\usepackage{algorithmic}
\usepackage{graphicx}
\usepackage{textcomp}
\usepackage{xcolor}
\usepackage[hyphens]{url}
\usepackage{fancyhdr}
\usepackage{hyperref}

\usepackage{caption}
\usepackage{subcaption}
\usepackage[revision]{revdiff}
\newcommand{\revise}[2]{\textcolor{black}{#1}}
\newcommand{\ignore}[1]{}
\newcommand{\bmc}{\emph{BMC\/}}

\newcommand{\todo}[2]{\textcolor{black}{#1}}
\long\def\hpcarevise#1{\textcolor{black}{#1}} 
\newcommand{\new}[2]{\textcolor{black}{#1}}
\newcommand{\newarun}[2]{\textcolor{black}{#1}}
\newcommand{\changearun}[1]{{\textcolor{black}{#1}}}
\usepackage{booktabs}
\usepackage{siunitx}
\usepackage{xcolor}
\usepackage{listings}
\usepackage{graphicx}
\usepackage{colortbl}

\usepackage{cite}
\usepackage{amsmath,amssymb,amsfonts}
\usepackage{algorithmic}
\usepackage{graphicx}
\usepackage{textcomp}
\usepackage{xcolor}
\usepackage[hyphens]{url}

\def\BibTeX{{\rm B\kern-.05em{\sc i\kern-.025em b}\kern-.08em
    T\kern-.1667em\lower.7ex\hbox{E}\kern-.125emX}}
\begin{document}

\pdfpagewidth=8.5in
\pdfpageheight=11in

\newcommand{\iscasubmissionnumber}{2385}

\pagenumbering{arabic}

\title{Striking the Right Balance between Compute and Copy: Improving LLM Inferencing Under Speculative Decoding}
\author{
\IEEEauthorblockN{Arun Ramachandran\IEEEauthorrefmark{1},
R. Govindarajan\IEEEauthorrefmark{2},
Murali Annavaram\IEEEauthorrefmark{3},
Prakash Raghavendra\IEEEauthorrefmark{1}\\
Hossein Entezari Zarch\IEEEauthorrefmark{3},
Lei Gao\IEEEauthorrefmark{3},
Chaoyi Jiang\IEEEauthorrefmark{3}}
\IEEEauthorblockA{\IEEEauthorrefmark{1}Advanced Micro Devices (AMD)\\
Email: \{aruncoimbatore.ramachandran, prakash.raghavendra\}@amd.com}
\IEEEauthorblockA{\IEEEauthorrefmark{2}Indian Institute of Science (IISc)\\
Email: govind@iisc.ac.in}
\IEEEauthorblockA{\IEEEauthorrefmark{3}University of Southern California (USC)\\
Email: \{annavara, entezari, leig, chaoyij\}@usc.edu}
}


\maketitle
\thispagestyle{plain}
\pagestyle{plain}





\begin{abstract}
With the skyrocketing costs of GPUs and their virtual instances in the cloud, 
there is a significant desire to use  CPUs for large language model (LLM) inference. 
KV cache update, often implemented as allocation, copying, and in-place strided update
for each generated token,   incurs significant overhead.
As the sequence length increases, the allocation and copy overheads dominate the performance. 
Alternate approaches may allocate large KV tensors upfront to enable in-place updates, but these matrices  (with zero-padded rows) cause redundant computations. 

In this work, we propose a new KV cache allocation mechanism called \emph{Balancing Memory and Compute} ({\bmc}). 
{\bmc} allocates, once every $r$ iterations, KV tensors with $r$ redundant rows, allowing
in-place update without copy overhead for those iterations, but at the expense of a small amount of redundant computation. 
Second, we make an interesting observation that the extra rows allocated in the KV tensors and the resulting redundant computation can be repurposed for  Speculative Decoding (SD) that improves token generation efficiency.
Last, {\bmc} represents a spectrum of design points with different values of $r$.  To identify the best-performing design point(s),
we derive a simple 
analytical model for {\bmc}. 

The proposed {\bmc} method achieves  an average throughput acceleration of up to \new{3.2$\times$}{} over baseline HuggingFace (without SD). \rnew{Importantly when we apply {\bmc} with SD, 
it results in an additional speedup of up to 1.39$\times$, over and above the speedup offered by SD.}
Further, {\bmc} achieves a throughput acceleration of up to \new{1.36$\times$}{} and \new{2.29$\times$}{} over state-of-the-art inference servers vLLM and DeepSpeed, respectively. 
\rnew{Although the {\bmc} technique is evaluated extensively across different classes of CPUs (desktop and server class), we also evaluate the scheme with GPUs and  demonstrate that it works well for GPUs.}

\end{abstract}


\maketitle

\section{Introduction}\label{sec:intro}
LLMs have been deployed for a variety of tasks such as chatbots and virtual assistants~\cite{amazon_codewhisperer,anthropic_claude,bing_ai}.  Improving their execution efficiency is key to making them affordable to the broader societal needs. But LLM inference latency is hobbled by large memory management overheads for the KV (key-value) cache, which is a constantly expanding data structure.   
In this paper, we focus on the LLM decode phase, which consists of
$L$ transformer layers, and within each layer, there is an attention block 
and a multilayer perceptron (MLP) computation.  
The attention block computes self attention using Key ($K$) and Value ($V$)
tensors which are of dimension $[B,L,n,D]$, 
where $B$ is the batch size, $D$ is the embedding or hidden dimension size,  $n$  is the number of tokens generated so far (including the initial prompt), and $L$ is the total number of layers in the model\footnote{Some literature reports the dimension as $[L,B,n,D]$, with the layer dimension ($L$) as the first dimension, since the tensors are stored separately for each layer.}. 
The computation of self attention is performed as a Scaled Dot Product Attention (SDPA).  SDPA involves two matrix multiplications: (i) Query ($Q$) with Key Transpose, which is scaled to get the Attention scores and (ii) Attention scores with Value matrix. With a batch size greater than 1 or with multiple heads, self-attention computation can be performed efficiently as a batch matrix multiplication (BMM).
\ignore {The efficient computation of BMM necessitates leveraging  the optimized BLAS libraries which, in turn, requires the $K$ and $V$ tensors to be contiguous in the virtual address space~\cite{vattention}.}
\revise{To avoid 
performance drops and maintenance burdens of custom kernels, efficient BMM computation demands that the Key and Value tensors occupy contiguous blocks of virtual memory ~\cite{vattention}.}


Due to the auto-regressive nature of the decode stage, the size of $K$ and $V$ tensors increases by one column for each iteration. 
Typically, this growth in memory size is handled by allocating larger $K$ and $V$ tensors of dimension $[B,L,(n+1),D]$ after each token generation. 
This approach, referred to as the \emph{iterative allocation} approach, was followed in HuggingFace's implementation~\cite{huggingface_transformers}.
The self-attention computation for the $(n+1)$-th token depends on 
 values of $K$ and $V$ calculated in the previous $n$ iterations. 
 To avoid recomputing these tensors, 
the Key-Value (KV) cache mechanism~\cite{kvcache} was proposed which copies the $K$ and $V$ tensors of the $n$-th iteration in the respective tensors of the current iteration in a strided manner to account for the increase in the second last dimension. 
This operation, referred to as the \emph{KV cache update}, involves two major tasks, allocating memory for the $K$ and $V$ tensors in each iteration and copying the $K$ and $V$ tensors.  
These two tasks incur significant overhead in the attention block computation.

To overcome the above overheads, one could propose an \emph{upfront allocation}, which allocates the $K$ and $V$ tensors once for the maximum context length ($N$).  The unused elements (corresponding to future tokens) are initialized to~0. The $K$ and $V$ vectors computed in each iteration are written in the appropriate location in the larger $K$ and $V$ tensors. \revise{Static \emph{upfront} memory allocation might create internal fragmentation limiting batchsize and throughput as noted in ~\cite{vattention}. The other}, challenge with upfront allocation is whether to perform matrix multiplications on the entire KV cache, including the dummy values, or to selectively use only the valid values in the KV cache. Selective use of only valid values requires  
\ignore {developing custom}  BLAS \rold{GeMM} kernel with strided access (to filter out the dummy zero values). Our experimental results (see Section~\ref{sec:strided_BMM}) show that the SDPA computation is up to 14.4$\times$ slower with strided access,  compared to performing GeMV operations on a full matrix of $K$ and $V$ tensors, including the computations on padded zeros. 




\noindent \textbf{Contribution \#1:} Given these tradeoffs, we make the case for using a hybrid approach that straddles between iterative and upfront allocations, \revise{ensuring contiguity in virtual  address space.},  More specifically, we design a new memory allocation scheme, called \emph{Balancing Memory and Compute} or {\bmc}, which enables iterative allocation, once every $r$ iterations, of $K$ and $V$ tensors having $r$ additional rows and performing a limited number of wasteful computations (of at most $r$ elements) in the next $(r-1)$ iterations.  Thus it balances the overheads due to wasteful computation with memory allocation and copying operations, \revise{while continuing to leverage the vast set of optimized libraries \changearun{enabled out-of-box.}}.  

\noindent \textbf{Contribution \#2:} This approach of allocating $r$ redundant memory rows can still lead to wasteful computations, albeit much less than upfront allocations. We then make an interesting observation to exploit these wasteful redundant rows: Most LLMs today use Speculative Decoding (SD)~\cite{leviathan2023fast, li2024eagle, chen2024sequoia}. SD is a technique that significantly accelerates LLM inference latency without sacrificing output quality, and with minimal resource overhead. SD leverages one or more smaller draft models to generate predictions efficiently for future tokens. Given that {\bmc} allocates $r$ redundant rows we make the case for repurposing these $r$ redundant rows to temporarily store the speculatively decoded K and V tensors thereby significantly eliminating padded zero computations. 

    
As SD verifies the speculated tokens together,
the computational pattern associated with SDPA changes from General Matrix-Vector Multiplication (GeMV) in auto-regressive decoding to General Matrix-Matrix Multiplication (GeMM) in SD. 
Thus combining {\bmc} with SD makes the  SDPA computation (using BLAS routines) more efficient, but this 
requires the speculatively decoded K and V  tensors to be placed in the virtual address space contiguously. Our approach in fact achieves this with the redundant space allocation.

\noindent \textbf{Contribution \#3:}  The {\bmc} approach provides a spectrum of design points by varying $r$.  To reduce the burden on the designer to select the appropriate $r$ we derive an analytical model that helps to identify the best-performing design points. An interesting feature of the analytical model is that the derived best-performing design point is dependent only on the maximum context length ($N$) and is independent of the LLM and its model parameters.

\noindent \textbf{Contribution \#4:} When performing any decoding with padded zeros,
the attention calculation, which involves a softmax operation, can introduce inaccuracies. To address this, 
we propose the use of a bias mask to avoid any additional computation overhead.

\ignore{ 
We conduct our evaluation on a wide range of systems, including AMD\footnote{AMD, the AMD Arrow logo, Genoa, Milan, Ryzen, and Instinct MI210, which are trademarks of Advanced Micro Devices, Inc. Other product names used in this publication are for identification purposes only and may be trademarks of their respective companies. @2025 Advanced Micro Devices, Inc. All rights reserved.}  Ryzen$^{TM}$ clients, 
datacenter servers with AMD Genoa and Intel Sapphire Rapids processors. 
{\bmc} consistently outperforms state-of-the-art inference engines  vLLM~\cite{Paged_Attention} and DeepSpeed~\cite{2022deepspeed}.  {\bmc} with SD achieves an additional improvement of up to $21\%$ over baseline Speculative Decoding without our proposed optimizations.
}


\noindent \textbf{Contribution \#5:}  We implemented {\bmc} in an end-to-end inference setup. Experimental evaluation demonstrates:
\begin{enumerate}
    \item {\bmc} achieves an average (geomean) speedup of \new{$3.2\times$}{} over baseline HuggingFace (without SD) \revise{on a wide range of desktop and server category systems}.
\item {\bmc} with SD achieves a speedup of $1.39\times$ over SD. 
\item {\bmc} implemented as an inference server, which supports multiple instance parallelism, outperforms state-of-the-art inference engines vLLM~\cite{Paged_Attention} and DeepSpeed~\cite{2022deepspeed} by \new{1.36$\times$}{} and \new{2.29$\times$}{}, respectively.
\item \revise{Our {\bmc} approach, the analytical model and its performance advantage continue to hold in GPUs as well, resulting in 1.4x~--~1.7x performance improvement over the baseline.}{Extensive experimentation on a wide range of systems including AMD\footnote{AMD, the AMD Arrow logo, Genoa, Milan, Ryzen, Instinct MI210  are trademarks of Advanced Micro Devices, Inc. Other product names used in this publication are for identification purposes only and may be trademarks of their respective companies. @2025 Advanced Micro Devices, Inc. All rights reserved.}  Ryzen$^{TM}$ clients, 
datacenter servers with AMD Genoa and Intel Sapphire Rapids processors, and on MI210 GPUs demonstrate that our BMC approach and the analytical model work equally well across all of them.}
\end{enumerate}

\subsection{Importance of CPU LLM Inference}
In this work we evaluated  {\bmc} extensively on CPUs, while also providing several key results on GPUs. We motivate the importance of studying CPU-based LLM inference here.  As the cost and critically the availability of GPUs hit a significant hurdle there is a growing demand \revise{for accelerating LLM inference on CPUs for a variety of end-user tasks on different platforms, "e.g.," the AI Pcs~\cite{ai-pc-opportunity}.}{to support LLM inference capabilities  on CPUs. For example,  AI PCs~\cite{ai-pc-opportunity} that improve productivity perform significant number of AI inferences on a CPU. Thus there is a growing demand for accelerating LLM inference on CPUs for a variety of end-user tasks. } 
The increase in CPU core counts, large configurable DRAMs, and support for a wide range of data precision formats 
enable LLMs to run efficiently on commodity CPUs, offering good performance per dollar~\cite{SLIDE, shen2023efficient, he2024inference}. 
For example, processor families such as AMD EPYC\texttrademark{} Genoa and Turin leverage VNNI (Vector Neural Network Instructions) to accelerate BLAS operations that are critical for LLM inference improvements. Similarly, 4th Gen Intel\textregistered{} Xeon\textregistered{} processors Sapphire Rapids are available for LLM low-cost deployments with EC2 Instances, and Granite Rapids support AMX (Advanced Matrix Extensions) with a throughput of 784 INT8 TOPS. These servers offer 500 GB/s of memory bandwidth, extendable up to 7 TB/s on Microsoft Azure HBv5 instances, enabling high-performance LLM inference. 

\revise{As reported in~\cite{idtechex_report}, CPU-based LLM inference has a growing market share (12--15\% share of the \$300--\$400B AI chip market). Also, evaluation  by industry leaders (Oracle~\cite{oracle_blog}, Cisco~\cite{cisco_documentation}, Dell~\cite{dell_infohub}, Intel~\cite{intel_amx}, HuggingFace~\cite{huggingface_blog}), recent research advances 
(NoMAD-Attention~\cite{nomad_attention}, FlexInfer~\cite{flexinfer}, Challenging GPU Dominance~\cite{challenging_gpu_dominance}, Dovetail~\cite{dovetail}, DeepSparse~\cite{deepsparse}, INT4 quantization on CPUs ~\cite{Efficientcpu}), 
and robust support on major cloud platforms like AWS EC2~\cite{intel_amx}, Google Cloud~\cite{larabel2023googlecloud} and Azure HBv4~\cite{azure_hbv4} all highlight the  cost-effectiveness, efficiency, scalability and accessibility of CPU inference.} {}
\section{Background} \label{sec:MHA}
\subsection {LLM inference with Multi Head Attention}

An LLM consists of a sequence of $L$ layers, each layer comprising Attention and Multi-Layer Perceptron (MLP) Blocks.  The attention computation starts with the current token $x$, and  computes the query, key and value vectors as:
\[ q = W_q{\cdot}x; ~~k = W_k{\cdot}x; ~~v = W_v{\cdot}x \]
where $W_q$, $W_k$, and $W_v$ are weight matrices.  
We refer to this as \emph{KV compute} task. 

The $k$ and $v$ vectors are concatenated to the $K$ and $V$ matrices from the previous iteration to produce the new $K$ and $V$ matrices. Thus for each iteration the size of $K$ and $V$ matrices grows by one vector. As mentioned earlier, this step involves allocating memory for the new $K$ and $V$ matrices and copying the values from the previous matrices. We refer to this task as \emph{KV cache update}.

The attention computation in each layer performs:
\[ Attention = softmax \left( (Q\cdot K^{T})/{\sqrt D} \right) \cdot V \]
where $Q$ represents the query vector, $K^{T}$ represents the transpose of the key matrix and $V$ corresponds to the value matrix.  The self-attention computation is performed as a Scaled Dot Product Attention (SDPA).
For performing the attention operation for a batch size $B$ and across $L$ layers, then the $K$ and $V$ matrices become tensors of dimension $[B,L,n,D]$, where $n$ is the current context length (this includes the prompt length along with the total tokens generated so far) and $D$ is the total number of hidden dimensions.  For a maximum context length of $N$, these tensors can be as large as $[B,L,N,D]$. 
For batch size $B=64$,  number of layers  $L=64$, hidden size $D=4096$, and a maximum context length $N=2048$,  the $K$ and $V$ tensors could require  137~GB with half precision.
In computing self-attention, a significant part of the $K$ and $V$ matrices (corresponding to previously generated tokens and their attention values) are recomputed for each new token generation.  To avoid this recomputation, a structure called \emph{KV Cache} stores the $K$ and $V$ matrices computed in the previous iteration and copies them into the new $K$ and $V$ matrices (size increased by one additional row or column) in the respective positions corresponding to the previously generated tokens. This avoids the need to recalculate information that has already been computed.  This optimization, known as KV cache optimization, is performed in all LLM implementations. 


\lstset{
    basicstyle=\tiny\ttfamily, 
    numbers=none, 
    keywordstyle=\color{blue},
    commentstyle=\color{green!60!black},
    stringstyle=\color{red},
    frame=single, 
}

\begin{figure}[h]
    \centering
    \begin{lstlisting}[language=Python]
 1 def update_cache(cache, sequence_no, val):
 2     dimensions = get_dims(cache)
 3     new_cache = allocate(sequence_no, dimensions)
 4     new_cache = cache.concat(val)
 5     return new_cache

 6 def KV_compute(x, wq, wk, wv):
 7     Q = x @ wq
 8     K = x @ wk
 9     V = x @ wv
10     return Q, K, V

11 def Scaled_dot_product_attention(Q, new_k_cache, new_v_cache):
12     attention_weights = BMM(Q, new_k_cache.transpose())
13     attention_prob = SoftMax(attention_weights)
14     attention_output = BMM(attention_prob, new_v_cache)
15     return attention_output

16 def Attention_block(x, k_cache, v_cache, sequence_no, wq, wk, wv):
17     Q, K, V = KV_compute(x, wq, wk, wv)                           #KV Compute
18     new_k_cache = update_cache(k_cache, sequence_no, K)           #KV Update
19     new_v_cache = update_cache(v_cache, sequence_no, V)
20     O = Scaled_dot_product_attention(Q, new_k_cache, new_v_cache) #SDPA 
21     return O
    \end{lstlisting}
    \caption{Attention Computation with Iterative Allocation.}
    \label{AttnBlkListing}
\end{figure}



\subsection{Attention Computation With Iterative Allocation:}

Figure~\ref{AttnBlkListing} shows the pseudo-code of the attention block with iteratively allocated KV cache implementation.  Note that in this implementation, a new $K$ and $V$ matrix is allocated in each iteration of token generation and the contents of the KV cache are copied into these matrices which, in the current iteration, becomes the KV cache, after the newly computed. $k$ and $v$ vectors are concatenated. We refer to this approach as \emph{iterative allocation}.  We identify two regions, (1) the \emph{KV cache update} region that includes copying of the  $K$ and $V$ cache to the newly allocated $K$ and $V$ matrices and concatenating the new row/column,  and (2) the attention computation region, referred to as \emph{SDPA} region.   
The number of elements that are copied when the $n$th token is generated is $B{\cdot}n{\cdot}D$ in each layer.  Thus, the total copying done across $L$ layers, for both $K$ and $V$ matrices, and for the generation of the maximum number of tokens $N$ is:\\
\begin{small}
\[ \sum_{n=1}^N 2{\cdot}B{\cdot}L{\cdot}n{\cdot}D  = 2{\cdot}B{\cdot}L{\cdot}N{\cdot}\frac{(N+1)}{2}{\cdot}D = B{\cdot}L{\cdot}N{\cdot}(N+1){\cdot}D \]
\end{small}

The computation FLOPs required for the self attention computation in the iterative allocation approach where the  $K$ and $V$
tensors are of exact size ("i.e.,", $[B,L,n,D]$) in each iteration, while $Q$ is $[B,1,D]$. 
The total FLOPs required for SDPA in iterative allocation is: \\
\begin{small}
    \[ \sum_{n=1}^N 2{\cdot}L{\cdot}B{\cdot}n{\cdot}D  = 2{\cdot}L{\cdot}B{\cdot}N{\cdot}\frac{(N+1)}{2}{\cdot}D =  
    L{\cdot}B{\cdot}N{\cdot}(N+1){\cdot}D \]
\end{small}

LLM architectures typically use MultiHead Attention using $H$ heads each of dimension $d$, such that $D=H{\cdot}d$. 
The compute FLOPs associated with MultiHead Attention is essentially the same as Single Attention ~\cite{shazeer2019fast}. 
In Multihead Attention, the $Q$, $K$, and $V$ vectors 
are reshaped to merge the  $H$ heads with the batch size $B$. For each individual layer of $L$, the dimension of  $Q,K,V$ vector becomes $[B{\cdot}H,{1},{d}]$ while the previous $KV$ matrix is of dimension  $[B{\cdot}H,{n-1},{d}]$, where $n$ is the current context length (including the new token). The concatenation of KV matrix with vector results in  $[B{\cdot}H,{n},{d}]$.The batch matrix multiplication of $(Q{\cdot}{K^{T}})$ results in a squeezed tensor of shape $[{B}\cdot{H},{1},{n}]$. The second operation associated with Scaled Dot Product Attention (SDPA) is element-wise softmax, where the contracted tensor remains as $[B{\cdot}H,{1},{n}]$. In the final step, the softmax outputs are batch multiplied with $V$, resulting in $[B{\cdot}H,{1},{d}]$. 

\subsection{Upfront Allocation Approach}
The memory allocation and copy overheads  of iterative allocation can be overcome by
an alternative strategy, referred to as \emph{upfront allocation}.  It allocates the $K$ and $V$ matrices corresponding to the maximum context length ($N$) upfront.  When the $n$th token is generated, the elements corresponding to tokens $(n+1)$ to $N$ of the matrices $K$ and $V$ are set to 0.  The newly computed $K$ and $V$ vectors in each iteration of token generation are written in place in this larger KV cache.  For the attention computation, the batch matrix multiplication is done with this larger $K$ and $V$ matrices. This approach avoids both the allocation and the copying overheads involved in the iterative approach.   However, there are two issues with this approach. 

First, the zero padded elements increase the amount of computation performed by the batch matrix multiplication. The total computation required with this approach is $2{\cdot}B{\cdot}N{\cdot}D$ (considering the two batch matrix multiplications involved in the self-attention computation). The total FLOPs\footnote{Note that FLOPs stands for the  Floating Point Operations (count), while FLOPS represents the rate, Floating Point Operations Per Second. Further, we consider multiply-and-accumulate (MAC) as a single operation.} required across $L$ layers,  for the generation of tokens $N$, is:\\ 
\[ \sum_{n=1}^N 2{\cdot}L{\cdot}B{\cdot}N{\cdot}D  =  
2L{\cdot}B{\cdot}N{\cdot}(N){\cdot}D \]
The computation requirement in the upfront allocation is roughly 2x compared to the iterative allocation approach.
The second problem with the 0-padded rows (and columns) is that they introduce spurious values in the softmax computation. The exponentiation performed in the softmax computation for each padded 0 impacts the softmax scores as $e^0=1$.  This affects the attention values and hence the accuracy of the LLM. 

One may consider a more selective matrix-vector computation by only using non-padded matrix rows. However, such an approach requires custom kernels that perform computations using strided accesses on a row-by-row basis. But as we show later this approach is significantly slower than performing even needless computations on padded zeros. 

\rnew{
\subsection {LLM Inference with Speculative Decoding }

Since LLMs generate tokens in an auto-regressive manner, each iteration has to wait for the prior iteration token output. To reduce this serialization SD is employed. The target large model works in tandem with one or more smaller (and hence faster) ``draft" models.  The draft model generates a sequence of $t$ speculative tokens, which are then verified in parallel by the target model.  Consequently, the query vector used by the original model transitions from a shape of $[B \cdot H, 1, d]$ to $[B \cdot H, t, d]$ to incorporate these tokens, effectively transforming GEMV operations into GEMM operations. 
If $m$ tokens are accepted (where $m \leq t$), the remaining unaccepted tokens are discarded, and only the $m$ accepted entries are used to update the KV cache during each iteration. 
SD has the potential to accelerate the throughput by a factor of $m$, where $m$ is the average number of accepted tokens across the sequence generation.
}


\section{Related Works}

\revise{
There have been several existing approaches for improving the efficiency of KV Cache management in LLM decoding.
Huggingface~\cite{huggingface_transformers} and DeepSpeed~\cite{2022deepspeed} implementations follow the iterative allocation approach and hence incur memory allocation and copying overheads.
The Paged Attention mechanism~\cite{Paged_Attention},  proposed to specifically handle the memory capacity issues in GPU,  allocates a chunk/block of virtual memory for a fixed number of tokens.
vLLM~\cite{vllmpagedattentionconcepts}, an inference serving solution that implements Paged Attention, performs this periodic allocation once every 16 or 32 tokens.   The periodic allocation of KV cache is the only commonality that {\bmc} shares with  vLLM. }{}

\revise{
The key differences between the two approaches are: (1) The periodic allocation Paged Attention results in \emph{non-contiguity} in the virtual address space. The non-contiguity hinders leveraging optimized BLAS libraries and out of box support with 
Scaled Dot Product Attention (SDPA) or FlashAttention~\cite{vattention}. Further, each allocation in vLLM implementation~\cite{vllmpagedattentionconcepts} is only for a partial token context~---
allocations for  different inputs of a batch ($B$) and for different heads ($h$) in a layer are done separately, increasing the extent of non-contiguity, and the overheads invoking BLAS kernels repeatedly even further. (2)~vLLM advocates a fixed block size of 16 or 32 \changearun{and tile sizes (determined by block and head size (model dependent)) that could be suboptimal for effectively leveraging the cache hierarchy.}   In contrast, {\bmc} identifies the best-performing allocation size through an analytical model \changearun { and optimal tile sizes determined by BLAS}.  (3)~In Paged Attention, the allocated  virtual blocks are used across token generation  to avoid KV cache copying completely, but at the expense of non-contiguity. {\bmc} trades off  minimal KV cache copy overhead (incurred once every $r$ iterations) to take advantage of optimized kernels and software compatibility. Despite the copying overhead, as shown in our quantitative comparison (see Section~\ref{comp-vllm-deepspeed}), {\bmc} outperforms vLLM. This corroborates well with the statement from  vLLM~\cite{Paged_Attention}  (in Section 7.1) that concedes its attention computation is 20–26\% slower than non-paged kernels due to block-table lookups and branching.}{}

\newarun{
  vAttention promotes contiguous layouts but has several practical constraints: (1) it depends on CUDA virtual memory APIs, tying compatibility to specific CUDA versions and excluding non‑CUDA backends ("e.g.," ROCm) and limits experimentation with CPUs, older CUDA devices and other accelerators; (2)  Vattention implicitly assumes small incremental KV growth. Each on‑demand physical allocation incurs overhead and  larger batch sizes or hidden dimensions increases raise end‑to‑end latency; (3) page granularity in the KB–MB range increases the risk of internal fragmentation (as noted by the authors); (4) it requires KV cache sizes to be fixed a priori, hindering dynamic techniques such as speculative decoding; and (5) current software integrations are largely confined to FlashInfer/FlashAttention variants. In contrast, BMC avoids virtual memory APIs, simplifies memory management, minimizes allocation frequency, adapts to dynamic techniques, and can integrate with a wide range of attention implementations and backends.
}{}
\rold{Compared to approaches like Paged Attention~\cite{Paged_Attention},  that use non-contiguous, virtual memory, 
our scheme offers an improved allocation policy.   It differs from methods like ORCA~\cite{orca} and vAttention~\cite{vattention} that use upfront and iterative allocations. 
}

Beyond KV cache management, different schemes for  efficient large language models have been proposed in the recent past. 
Of particular interest to this paper are inference frameworks, 
"e.g.," DeepSpeed~\cite{2022deepspeed}, Megatron~\cite{megatron}, OpenLLM~\cite{pham2023openllm}, vLLM~\cite{Paged_Attention}, and TensorRT-LLM~\cite{nvidia2023tensorrt}. 

DeepSpeed~\cite{2022deepspeed} optimizes large models for distributed inference and training across multi-node clusters to reduce communication overhead and memory usage. Megatron~\cite{megatron} implements model parallelism on MLP layers with parameter splits.   
Flash Attention~\cite{dao2022flashattention} accelerates attention operations on GPUs with efficient tiling. 
Approaches like Flexgen~\cite{flexgen} utilize efficient batching strategies to minimize weight offloading to and from disk and improve the throughput of generative inference on a single GPU. 
Flash Attention-2~\cite{dao2023flashattention2} improves upon the prior version with efficient parallelization and workload distribution. 
Flash Attention-3~\cite{shah2021flashattention}  overlaps compute and data movement with block-level quantization to further enhance the LLM performance.  

KIVI~\cite{liu2024kivi} and KVQuant~\cite{hooper2024kvquant} are quantization approaches to  reduce the KV cache size. Streaming LLM~\cite{xiao2023streamingllm} and Heavy-Hitter Oracle (H2O)~\cite{zhang2023h2o} \revise{are token-pruning methods that limit the KV cache size by carefully selecting a fixed number of recent and performance-critical tokens.}{is a KV cache eviction strategy that retains a balance between recent and performance-critical tokens.}
In contrast our {\bmc}  approach stores the entire KV cache (with its full precision), and yet improves the performance of the attention block. 

Pruning or model compression is an orthogonal approach used to improve the inference speed of LLMs. A plethora of compression methods have been reviewed in~\cite{zhu2023survey}~\cite{tang2024transformersurvey}~\cite{wang2024compressionsurvey}~\cite{park2024comprehensivesurvey}. 
Our work is orthogonal to these and can coexist with them. 

\section{Motivation}

\begin{figure} [htb]
\centering	
    \includegraphics[width=6cm]{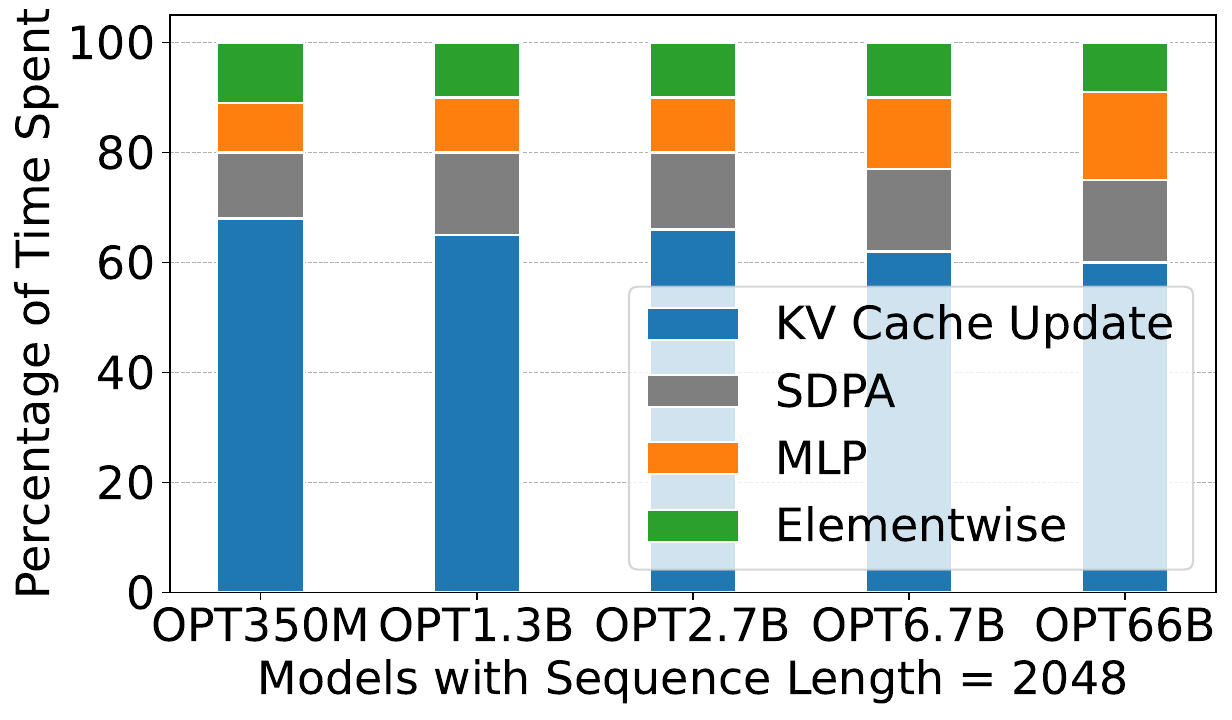}
	\caption{Execution Time Contribution for OPT Models.}
          \label{fig:ExecTimeContrib}
\end{figure}

We motivate the work presented in this paper by understanding how much each subtask
within LLM decoding contributes to the end-to-end inference time  on an AMD Genoa Server\footnote{Section~\ref{experimental_study} describes the experimental methodology and the platforms used.}. For a wide range of models ("e.g.," OPT-350M to OPT-66B) and with a maximum context length of 2048, the attention layer contributes roughly  75\% to 80\% of the inference time, as shown in Figure~\ref{fig:ExecTimeContrib}.  

\ignore{
\begin{figure} [htb]
\centering	
    \includegraphics[width=6cm]{Diagrams_pdf/Time_across_different_seq.pdf}
	\caption{Execution Time Contribution for Different Sequence Lengths.}
          \label{fig:ExecTimeContribSeq}
\end{figure}
}

Within the attention layer,  we find that the KV cache update 
contributes 60--70\% of the end-to-end inference time. The attention computation (marked as SDPA) itself takes 15-20\% of the time 
This indicates that considerable benefits can be achieved by optimizing the attention layer.  

\begin{figure} [htb]
\centering
	\includegraphics[width=6cm]{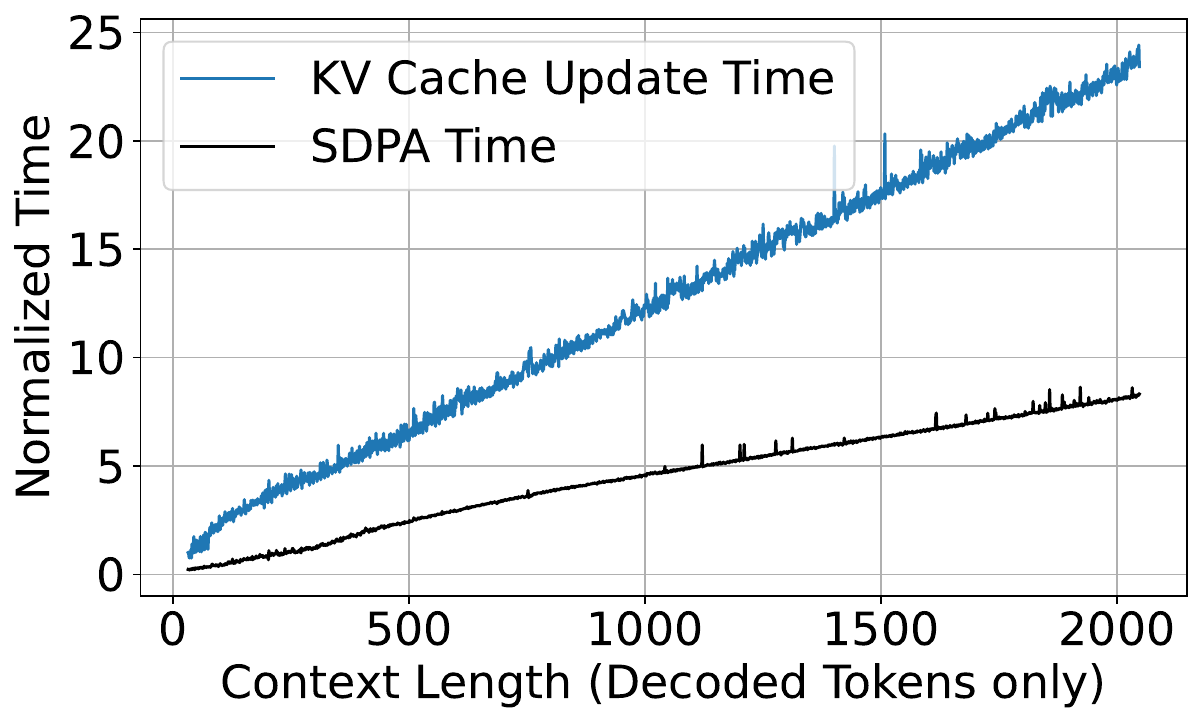} 
	\caption{ \rnew{KV Cache update  and SDPA Computation  Time under Iterative Allocation. }}
        \label{KVCacheupdate}
\end{figure}

Figure~\ref{KVCacheupdate}  presents the time taken by the KV cache update operation and the SDPA calculation for 1 layer of the OPT-2.7B model on a Genoa server as the current context length increases from 1 to 2048.  In this experiment, we have used a batch size of 512 with iterative allocation strategy ($K$ and $V$ matrices of appropriate size are allocated in each iteration). 
While the time taken for both operations increases with increasing context length,  KV cache update time grows significantly faster compared to the SDPA computation. 

Although this may appear somewhat surprising initially, our analysis in the previous section reveals that the number of operations required for KV cache update is roughly the same as the SDPA calculation ($O(B{\cdot}L{\cdot}N^2{\cdot}D)$). 
At equal number of operations, memory operations are slower than compute operations, as the  
memory speed in modern CPU servers is a few hundred GB/sec while the peak compute performance is of the order of a few TeraFLOPS.
This observation essentially means that the KV cache update is bandwidth-bound and is bottlenecked  for large context lengths~\cite{shazeer2019fast}.  

Next, we ask the question of what happens if upfront allocation is done with zero padded rows  for the KV cache and the SDPA is done with the padded zeroes. Note that this approach avoids repeated memory allocation of new $K$ and $V$ matrices and  copying  the KV cache values into them. For now we ignore the inaccuracies caused in the self-attention computation due to the 0-padding. We measure the computation time for the attention block which includes the computation of the attention values using the larger $K$ and $V$ matrices (corresponding to the maximum context length) in each iteration and their update in place in the KV matrix.  We compare the attention  computation time of the upfront approach with that of the iterative allocation approach.
\begin{figure} [htb]
\centering
	\includegraphics[width=6cm]{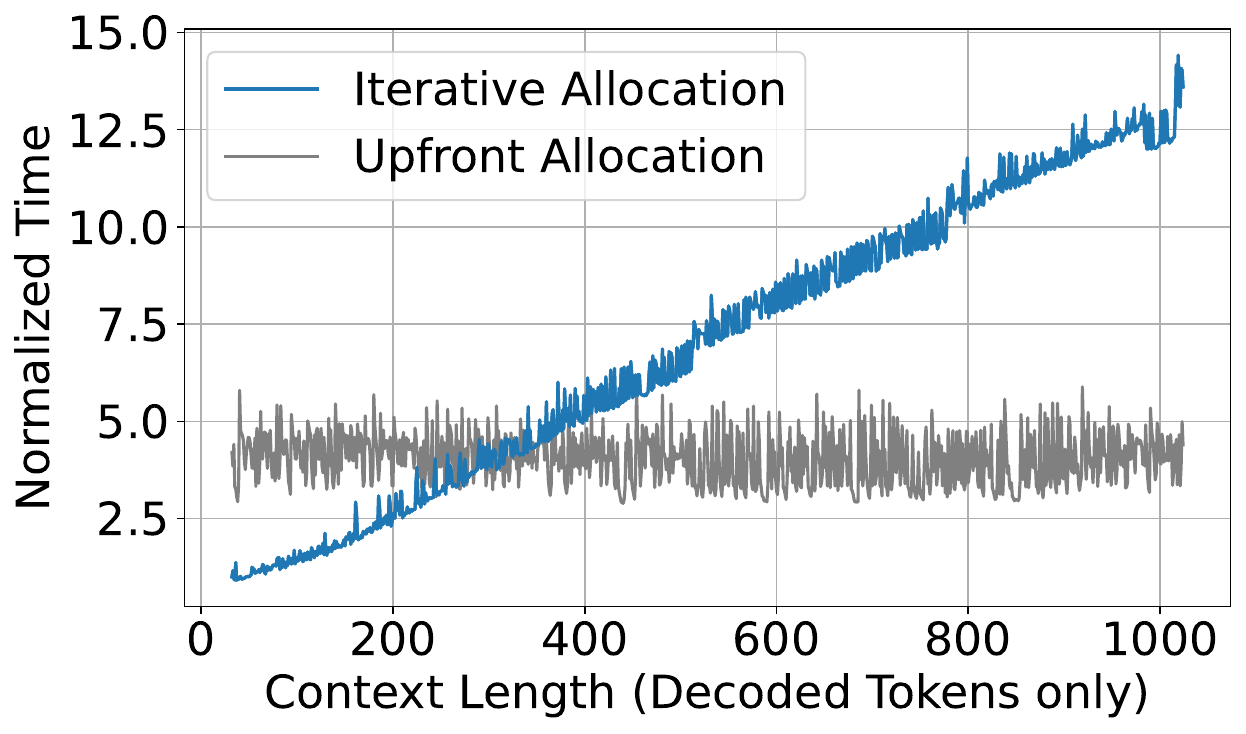}
	\caption{Total Time taken for Attention Layer -- Upfront versus. Iterative Allocation.}
        \label{upfrontAttentionComp}
    \end{figure}
Figure~\ref{upfrontAttentionComp} shows that while the attention block computation time increases with the context length for iterative allocation, it remains nearly constant (except for some small fluctuation) for the upfront approach. Despite wasteful computation (due to zero padding), the upfront allocation results in significantly lower time (lower by 1.9$\times$) than iterative allocation.

\section{{\bmc}: Balancing Memory and Compute}   \label{sec:bmc}
\ignore {
\section{{\bmc}: Balancing Memory and Copy}   \label{sec:bmc}
}
Motivated by the above observations we propose {\bmc} (\emph{balancing memory and compute}), a technique in which $K$ and $V$ matrices are allocated once every $r$ iterations, large enough to accommodate growth of the context by $r$-tokens.  In other words, the allocated $K$ and $V$ matrices would have up to $r$ zero-padded rows. In the next $r$ iterations, the same matrices are used for in-place updates of the newly generated and k and v tensors (hence avoiding any copying of elements). After the $r$-th iteration, new $K$ and $V$ matrices are allocated and the values from the previous $K$ and $V$ matrices are copied.  Thus we incur copy overhead once every $r$ iterations and wasteful computation for the next $(r-1)$ iterations.  Note that the wasteful computation is limited to at most $(r-1)$ rows. The value of $r$ relates to the number of allocations ($T$) of  the $K$ and $V$ tensors as in 
\( T = N/r, \)
where $N$ is the maximum context length. 

\begin{figure} [htb]
 \centering
	\includegraphics[height=3.0cm]{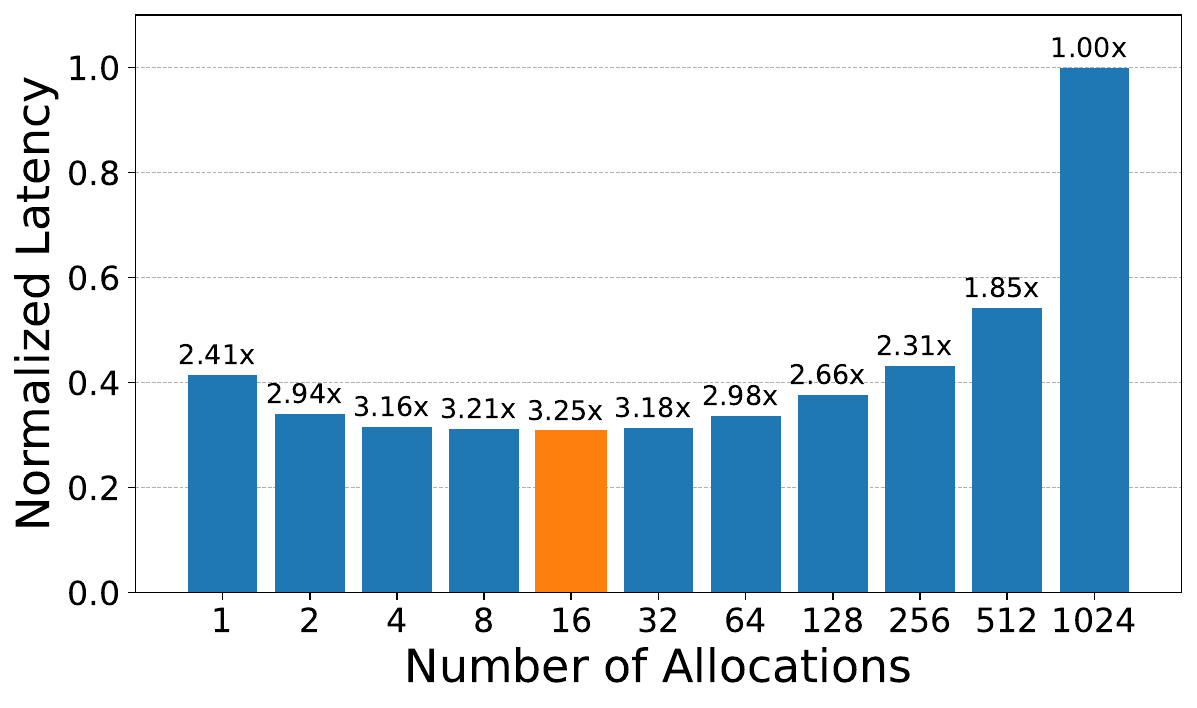}
	\caption{Speedup in Attention Block Latency with {\bmc}.}
        \label{SWEET_SPOT}
    \end{figure}

Figure~\ref{SWEET_SPOT} shows the attention block computation time for the OPT-13B model~\cite{vicuna} with a batch size of 8 and a maximum context length of 1024. The x-axis represents the number of allocations $T$. At $T=16$,  the attention computation time is the lowest and is \todo{3.25$\times$ and 1.34$\times$} lower than the iterative and upfront allocation approaches.  This tradeoff suggests that the inference time can be optimized by choosing an appropriate value for $T$. However, the choice of selecting $T$ must be automated to remove the burden of manual hyperparameter tuning. 




\subsection{Optimal Strategy} \label{OPT_STRATEGY}
We develop an analytical model for 
the total time of memory and compute operations\footnote{For simplicity, we derive the analytical expression for  attention computation with a single head.}. 
As the compute time for calculating the $K$ and $V$ vectors (KV compute time) in each iterations remain the same across all iterations, we do not consider it in our analytical model. 

For our {\bmc} approach, we express the context length $n$ (the number of tokens generated so far) in terms of $r$ as:
\[ n = i{\cdot}r+j,\] 
where $j~\in~[0,(r-1)]$,  $i$ varies from~1 to~$T=N/r$ and $N$ is the maximum context length. Without loss of generality, let us assume $r$ divides $N$ evenly.

In the {\bmc} approach, $K$ and $V$ are allocated, with $r$ additional zero-padded rows, once every $r$ iterations. That is, in iteration $i{\cdot}r$, new  tensors of dimension $[L,B,((i+1){\cdot}r),D]$ are allocated for $K$  and $V$
and the KV cache copy operation takes place.  
In the subsequent $(r-1)$ iterations, the same $K$ and $V$ matrices are used without any copying.  But these iterations perform redundant computation of up to $(r-1)$ rows.  

Let us now derive the expression for the KV cache update and attention computation in $(i{\cdot}r)$th iteration.
\rold{ We ignore the cost of memory allocation, which occurs once every $r$ iterations. }
 \ignore{However, as the memory locations of the tensor are accessed/updated, a number of page faults proportional to the size of the tensor would be incurred.  Assuming each entry of the matrix to be 2B (half precision), the number of bytes allocated for the $K$ and $V$ matrices is
\(4{\cdot}B{\cdot}L{\cdot}((i+1){\cdot}r){\cdot}D.\) 
Thus accessing/updating these elements would incur 
\begin{eqnarray}
    \mbox{KV Alloc.Time} = 4{\cdot}B{\cdot}L{\cdot}((i+1){\cdot}r){\cdot}D/p 
\end{eqnarray}
where $p$ is the page size.}
To derive the KV cache copy time, we assume a memory bandwidth (BW) and efficiency factor $\alpha$ for memory bandwidth. Then the time taken to perform the KV update is given by: 
\begin{eqnarray}
    \mbox{KV Cache Copy Time} & 
    = \frac{4{\cdot}B{\cdot}L{\cdot}((i+1){\cdot}r){\cdot}D}{(\alpha{\cdot}BW)}    
\label{eq;KVCacheUpdate}
\end{eqnarray}
\rnew{We assume a constant time $C_0$ for KV cache allocation. } 
It should be noted that both allocation and copying of KV cache happen only once every $r$ iterations in our {\bmc} approach.

Next we calculate the Scaled Dot-Product Attention (SDPA) time for the $(i{\cdot}r)$-th iteration.  For each request in a batch, it involves two matrix multiplications of dimension $[1, D]~\times~[D,((i+1){\cdot}r)]$ and $[1,((i+1){\cdot}r)]~\times~[((i+1){\cdot}r),D]$.  The number of MAC operations involved in the SDPA computation for a batch size of $B$ and for  $L$ layers is: $2~{\cdot}B{\cdot}L{\cdot}((i+1){\cdot}r){\cdot}D$.  If the target system has a computation capability of  $C$ FLOPS and achieves a compute efficiency\footnote{Compute efficiency is defined as the ratio of achieved FLOPS in a computation to the total peak FLOPS of the system.} of $\beta$, then the SDPA computation time is given by: 

\begin{small}
\begin{equation}
\mbox{SDPA Compute Time} = \frac{2~{\cdot}B{\cdot}L{\cdot}((i+1){\cdot}r){\cdot}D}{(\beta{\cdot}C)} 
\label{eq:SPDA}
\end{equation}
\end{small}
The SDPA computation happens in every iteration and the size of $K$ and $V$ matrices are unchanged during the iterations $i{\cdot}r$ to $(i+1){\cdot}r-1$.  Thus, the total computation time  (KV cache update time plus the SDPA computation time) for these $r$ iterations is given by:

{\small
\makebox[\linewidth][l]{%
\scalebox{0.9}{
\begin{minipage}{\linewidth}
\begin{align}
    \mbox{Time for $r$ iters.} = & \frac{4{\cdot}B{\cdot}L{\cdot}((i+1){\cdot}r){\cdot}D}{(\alpha{\cdot}BW)} + C_0 + 
    \frac{2{\cdot}r{\cdot}B{\cdot}L{\cdot}(((i+1){\cdot}r){\cdot}D)}{(\beta{\cdot}C)}
\label{eq:rIterTime}
\end{align}
\end{minipage}
}
}
}

\ignore{
 \begin{small}
 \begin{align}
          \mbox{Time for $r$ iters.} = & \frac{4{\cdot}B{\cdot}L{\cdot}((i+1){\cdot}r){\cdot}D}{(\alpha{\cdot}BW)} + C_0 + 
  \frac{2{\cdot}r{\cdot}B{\cdot}L{\cdot}(((i+1){\cdot}r){\cdot}D)}{(\beta{\cdot}C)}
\label{eq:rIterTime}
 \end{align}
\end{small}
 }
From the above equation, we can estimate the total time for generating the maximum number of tokens $N=T{\cdot}r$ as:
  \begin{small}
  \begin{align}
      \mbox{Time for $N$ iters.}  = &    
       \sum_{i=0}^{T-1} \left( \frac{4{\cdot}B{\cdot}L{\cdot}((i+1){\cdot}r){\cdot}D}{(\alpha{\cdot}BW)} \right) + T{\cdot}C_0 + \nonumber \\
 & \sum_{i=0}^{T-1} \left( \frac{2{\cdot}r{\cdot}B{\cdot}L{\cdot}(((i+1){\cdot}r){\cdot}D)}{(\beta{\cdot}C)} \right)
    \label{eq:NIterTime}
 \end{align}
  \end{small}
{\small
\makebox[\linewidth][l]{%
\scalebox{0.85}{
\begin{minipage}{\linewidth}
\begin{align}
      \mbox{Time for $N$ iters.}=&  
    \left( \frac{4{\cdot}C1{\cdot}T{\cdot}(T+1){\cdot}r}{(2{\cdot}\alpha{\cdot}BW)} \right) + T{\cdot}C_0 +  \nonumber  
    &   \left( \frac{2{\cdot}C1{\cdot}T{\cdot}(T+1){\cdot}r^2}{(2{\cdot}\beta{\cdot}C)} \right)
 \end{align}  
 \end{minipage}
}
}
}
\\ \\
 Here $C1=B{\cdot}L{\cdot}D$.  Replacing $r$ by $N/T$, we get: 
 \begin{small}
\begin{flushleft}
\scalebox{0.9}{
\begin{minipage}{\linewidth}
\begin{align}
~~ & \mbox{Time }  \mbox{for $N$ iters.} =  
     \left(\frac{2{\cdot}C1{\cdot}N{\cdot}(T+1)}{\alpha{\cdot}BW} \right)  
  +  T{\cdot}C_0+\left(\frac{C1{\cdot}N^2{\cdot}(1+\frac{1}{T})}{\beta{\cdot}C}\right) \nonumber \\
& = \left( \frac{2{\cdot}C1{\cdot}N{\cdot}T}{\alpha{\cdot}BW} \right) +  
    \left( \frac{2{\cdot}C1{\cdot}N}{\alpha{\cdot}BW} \right)  + T{\cdot}C_0
  +  \left( \frac{C1{\cdot}N^2}{\beta{\cdot}C} \right)
   + \left( \frac{C1{\cdot}{N^2}}{\beta{\cdot}C{\cdot}T} \right)  
   \label{eq:NiterFinal}
 \end{align}
 \end{minipage}
 }
 \end{flushleft}
  \end{small}

To find the value of $T$ that results in minimum time for the computing the attention block, differentiate\footnote{We realize that the variable $T$ and the function are not continuous. However, for simplicity, we show the analysis with differentiation rather than with finite difference equation. A quick derivation with finite difference equation shows an equivalent result, but a with change in the constant value.} the RHS of Equation~\ref{eq:NiterFinal} w.r.t. $T$ and equate to 0.  
\begin{small}
   \begin{equation} {\label{Opt-T-eq}}
 \left( \frac{2{\cdot}C1{\cdot}N}{\alpha{\cdot}BW} \right) + C_0 - 
    \left( \frac{C1{\cdot}{N^2}}{\beta{\cdot}C{\cdot}T^2} \right) = 0 
\end{equation}
\end{small}
\revise{Our experiments in result section show memory allocation time is negligible with non iterative schemes. Hence $C_0$ can be ignored. }.
Solving the above equation gives:
\begin{small}
\begin{equation}
T = \sqrt{N \cdot \frac{\alpha \cdot BW}{2 \cdot \beta \cdot C}} \quad \text{or} \quad T \propto \sqrt{N}
\end{equation}
\end{small}

\ignore{
In the above equation, we have assumed that the values of $\alpha$ and $\beta$  are constant.  In practice, the efficiency of BMM ($\beta$) changes with the matrix size.  So is the memory bandwidth efficiency ($\alpha$).  We demonstrate empirically that the ratio $\alpha/\beta$, however, remain nearly constant. 
Figure~\ref{GEMM_KV_EFFICIENCY} presents the efficiency ratio of batch matrix multiplication operations to KV cache update operation, measured on AMD 9654 processor for the  OPT 6.7B model with batch size {$B=16$}  It remains nearly constant over a period of time. We expect the trend to hold true across models and systems. 

\begin{figure} [htb]
	\includegraphics[width=8cm , height = 4cm]{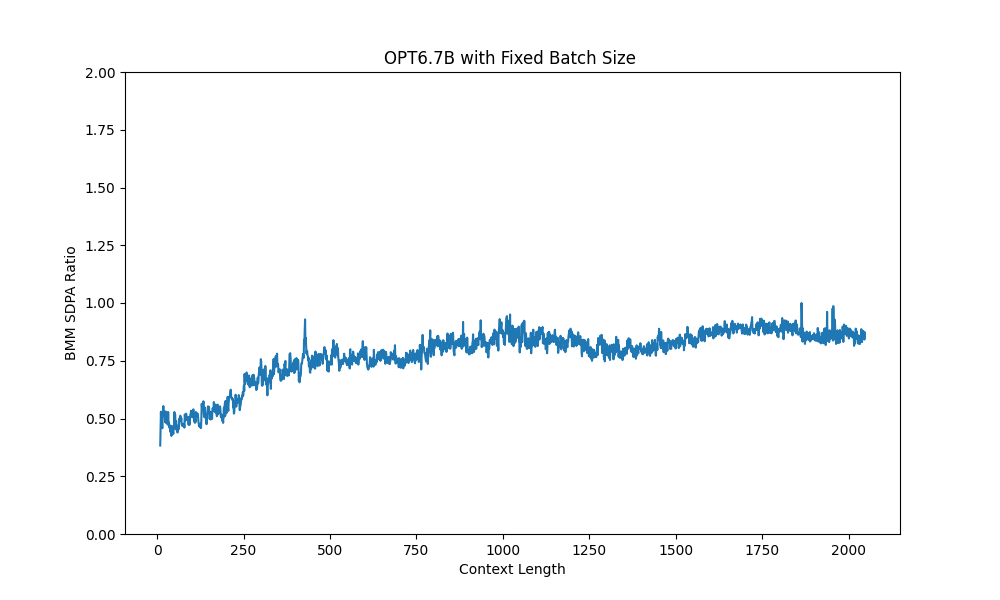}
	\caption{KVCache Scaled dot product efficiency ( Batch Size = 16) }
        \label{GEMM_KV_EFFICIENCY.}
    \end{figure}
\FloatBarrier

}

    
Our formulation indicates that the optimal number of allocations ($T$) is proportional to the square root of the maximum context length ($N$).  
Interestingly, $T$ is independent of the LLM  and model parameters. We demonstrate this behavior in our experimental evaluation. 

Before proceeding further, a few words of caution are in order. As remarked earlier,  the expression for the attention block compute time is not a continuous function. Also $T$ and $r$ can only take integer values. Further, choosing $T$ and $r$ as powers of 2 would result in higher efficiency for BMM. However, the above equation for optimal $T$  may result in non-integer values.  Hence we follow an approach where we compute the optimal value of $T$ using Equation~\ref{Opt-T-eq}, but round it to the nearest power of 2.
Since the approach approximates $T$, it  
may not guarantee the \emph{minimum} value for attention block execution  time. However,  
our experimental evaluations show that the value of $T$ obtained using our approach, in practice, does result in lower attention block execution time and the relationship between $T$ and $\sqrt{N}$ exists for the best performing design points. 
\ignore{
\rnew{ While our modeling does not include memory allocation and zero-padding costs in the analytical model, their effects can be folded into the constants C' in the equation. Specifically both memory allocation and zero padding costs are  observed to be proportional to the number of iterations, N. Including these costs in the model leads to an updated constant C', ensuring that the overall model remains consistent with the formulation. The optimal value for T is derived as:

\begin{equation}
  T = \sqrt{N \times C'}
\end{equation}

The correctness of the above analysis is also validated by our experimental evaluation.}
}

\subsubsection{Extension to Group Query Attention} \label{sec:GQAExtension}

\hpcarevise{
{\bmc} extends naturally to Group Query Attention (GQA). 
In GQA, the key and value projections are shared across multiple query heads. Hence the memory and compute costs lower by a factor of $G$ (number of groups).  Further with quantization, the size of KV tensors reduce by an additional factor of $Q$ (precision). Compressed KV heads are reused across query head groups during SDPA, improving cache locality and arithmetic intensity. While GQA reduces KV overheads by $GQ$, the SDPA operations remain unchanged due to the  reuse factor by $G$.}
\hpcarevise{
Thus the KV cache copy time for GQA with quantization is
\begin{eqnarray} \mbox{KV Cache Copy Time} = \frac{4{\cdot}B{\cdot}L{\cdot}((i+1){\cdot}r){\cdot}D}{(\alpha{\cdot}BW{\cdot}Q{\cdot}G)}
\label{eq:KVCacheUpdate_GQA} \end{eqnarray} 
This preserves the analytical formulation of the base case, scaled by the factor $G \cdot Q$, where $G$ is the number of groups and $Q$ is the quantization factor. 
}{}

\subsection{Accuracy with Padded Attention } \label{sec:MaskOps}
Next we address the issue of inaccuracies introduced  due to the zero padding 
in the softmax operation (which computes $e^0 = 1$).  To adjust this, the zeros introduced in the $Q{\cdot}K^T$ matrix, due to the zero-padded rows in the $K$ matrix, should be set to the smallest representable value in half-precision ($\approx -10^9$),
so that the softmax computation $e^{-10^9}$ results in a value which is closer~$0$.  

We propose to implement this using a mask which is applied to $Q{\cdot}K^T$, where $K$ is the key tensor for the maximum context length of $N$ and for a single layer.  Recall that $Q{\cdot}K^T$ is of dimension $[B,1,N]$. For this discussion, let us consider the attention computation performed as Multihead Attention (MHA). 
For MHA, $Q{\cdot}K^T$ is of dimension $[B{\cdot}H,1,N]$, where $H$ is the number of heads. Let us consider iteration $n=i{\cdot}r$ (iteration for generating the $(i{\cdot}r$th token in the decode phase).  Our allocation strategy allocates larger $K$ and $V$ matrices of dimension $[B{\cdot}H,((i+1){\cdot}r),d]$ that can hold values for the next $r$ iterations. Here the $Q{\cdot}K^T$ tensor is of dimension $[B{\cdot}H,1,((i+1){\cdot}r)]$
Thus the mask required should also be of dimension $[B{\cdot}H,1,((i+1){\cdot}r)]$.  Here, we set the first $i{\cdot}r + j $ columns to~$\mathbf{0}$ (a zero vector), and the remaining $r-j$ columns to contain $-10^9$ (a vector containing the smallest representable value in half-precision ($\approx -10^9$).  Note that the same mask can be reused across all the layers, and for the next $r$ iterations. Further, we can reduce the mask to the dimension $[1,((i+1){\cdot}r)]$  and broadcast the mask efficiently over the batched dimension $B{\cdot}H$.  We achieve this by using the mask as the bias in broadcast addition, which is supported by BLAS~\cite{bhaskaracharya2020automatic}. Thus the mask is converted into a bias operation and easily fused with General Matrix Multiplication, without incurring significant computation overhead. 

\section{BMC  with  Speculative Decoding}
\revise{{\bmc} scheme allocates $(r-1)$ additional zero-padded rows, once every $r$ iterations. This extra space can be repurposed for speculative decoding.}{{\bmc} strikes a balance between iterative and upfront allocation by limiting overhead to at most $(r-1)$ padded rows. However, further memory usage efficiency is possible in speculative decoding by reusing the $(r-1)$ padded rows. In particular, they can be used to verify candidate tokens without incurring additional memory allocation.} 


\subsection{Repurposing Zero-Padded Rows and Wasteful Compute}

\begin{figure} [htb]
    \centering
	\includegraphics[width=8cm]{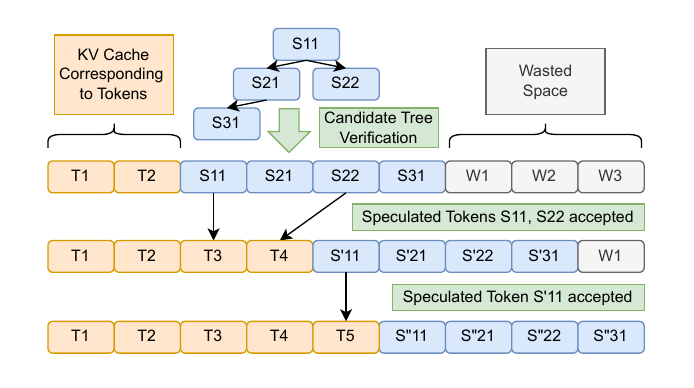}
	\caption{Repurposing Wasteful compute.}
        \label{fig:Redundant_Recompute}
    \end{figure}

While there are multiple approaches to SD, we follow the optimal tree topology~\cite{chen2024sequoia} for SD. We explain how we repurpose the padded rows using an illustration. Figure \ref{fig:Redundant_Recompute} illustrates a token tree speculated by the draft model, representing a candidate tree of size four. In this example, two candidate sequences are explored: (S11, S22), and (S11, S21, S31). Let us assume that {\bmc} has seven padded rows currently. The tree of four nodes (S11, S22, S21, S31) are allocated within the padded rows  to store speculative tokens for verification. This reuse reduces the number of wasted rows, "i.e.," rows that incur redundant SDPA computation, from seven to three in this example.
Let the sequence (S11, S22) be accepted as a valid contiguous sequence, leading to an update of the corresponding KV Cache. This updated cache is then 
used in the next iteration for next four speculative candidates (S$^\prime$11, S$^\prime$21, S$^\prime$22, S$^\prime$31,) generated by the draft model, further reducing redundant computation to 1. 
If the candidate S$^\prime$11 is accepted, then, in the subsequent iteration,  the speculative tokens would occupy the remaining zero-padded rows, completely eliminating the wasted computation.  
When the available zero-padded rows is less than $k$, the number of speculated tokens, our approach could either limit the number of speculated 
tokens (to the available rows) or do a new KV cache allocation for the required size.  We follow the former 
approach.

\ignore{ 
The optimal value for T  is derived as:

\begin{equation}
  T = \sqrt{N \times C'}
\end{equation}

In the case of speculative decoding, ~\cite{chen2024sequoia} investigates the speedup, which is proportional to 
\[  
\frac{G(k,d)}{t(k) + d \cdot c}.  
\]  

Here, G(k,d) — denoted as $M'$ — represents the average accepted length during speculative decoding (with K referring to the candidate set size and d to the maximum number of tokens that can be accepted). Meanwhile, t(k) denotes the time required to verify candidate sets, including the overhead associated with draft model speculation.
}

\subsection{Optimal Strategy with Speculative Decoding}
How does the optimal allocation strategy change for SD? Assuming the mean average accepted token is~$m$,  we note the following: 
\ignore{
\begin{enumerate}
\item The number of iterations is reduced from $N$ to $N'$ where $N' = \frac{N}{M'}$. ( $M'$ denotes the mean accepted length ) 
\item The KV Cache still requires update of all elements, so there are $N' \times M'$ total updates.
\item Although the parallel computation for SDPA increases by a factor of $K$, the overall SDPA compute workload decreases, as it now runs for only $N'$ iterations.
\end{enumerate}
}
\begin{enumerate}
\item The number of allocations $T$ remains the same. Hence the KV cache copying overhead remains the same as in the {\bmc} approach. 
\item The number of iterations between two consecutive KV cache allocations is reduced from $r$ to $r/m$.   During these iterations, the SDPA computation cost for verifying the $k$ speculated tokens increases by a factor of $k$, and the GeMV associated with SDPA becomes  GeMM operation. It is to be noted that verification of $k$ speculated tokens is a parallel computation and the computation efficiency ($\beta^\prime$) of GeMM call is higher.
\item We ignore the cost of  generation of speculative tokens, as they are smaller (using the draft model), and will also be incurred even in the base speculative decoding method.
\end{enumerate}
Using these, the modified time for $N$ iterations is written as:
 \begin{small}
   \begin{align}{\label{TimeNIter}}
 \mbox{Time  }  \mbox{for $N$ iters.} =  &
     \left( \frac{2{\cdot}C1{\cdot}N {\cdot}(T+1)}{\alpha{\cdot}BW} \right) + T{\cdot}C_0 + \nonumber \\
       & \left( \frac{C1{\cdot}k{\cdot}\frac{N^2}{m}{\cdot}(1+\frac{1}{T})}{\beta^\prime{\cdot}C} \right) 
 \end{align}
  \end{small}
Solving for optimal T results in 
\( T \propto \sqrt{\frac{N}{m}} \).

\section{Experimental Evaluation Methodology} \label{experimental_study}

We evaluate {\bmc} under both standard auto-regressive and speculative decoding. Experiments are run on a server with an AMD{\footnote{AMD, the AMD Arrow logo, AMD EPYC\textsuperscript{TM} (Genoa, Milan), AMD Ryzen\textsuperscript{TM}, AMD Instinct\textsuperscript{TM} MI210, AMD ROCm\textsuperscript{TM} are trademarks of Advanced Micro Devices, Inc. @2025 Advanced Micro Devices, Inc.} Genoa 9654 processor (96 physical cores, 768 GB DDR4 RAM). For ablation studies, we use an AMD Milan 7763 server (64 cores), an AMD Ryzen 7840 desktop (8 cores), and a server with a 60-core Intel Xeon Platinum 8490H. All server runs use cores and memory from a single socket, following standard practice. Accordingly, we set the number of threads to the number of physical cores per socket.

We run experiments using OPT models~\cite{huggingface_transformers}: 
OPT-
350M, 1.3B, 2.7B, 6.7B, and 66B. Since Speculative Decoding frameworks like SpecBench do not support OPT, we use LLaMA models, LLaMA-7B, 13B, and 70B, for those results. Although we conduct the full set of experiments, due to space limits, we sometimes show results for limited variants. Our findings remain consistent across all variants. We vary context lengths (512–2048 tokens) and batch sizes (1–128). Unless noted, results use a 2048-token context.

Before we discuss the performance results, we  demonstrate that our {\bmc} scheme does not result in any loss in accuracies. For this we use standard benchmarks such as Boolq and OpenQa~\cite{boolq,openqa}. along with the sample script and prompts from ~\cite{huggingface_transformers_issue_17653}.
 The perplexity scores and output tokens of the baseline and {\bmc} schemes match exactly.

 \section{Results}
 First, we validate the analytical model with experimental data. End-to-end performance of {\bmc} without and with speculative decoding are presented next. We then compare {\bmc} with state-of-the-art (SOTA) inference servers and provide an insight into the observed performance improvements. Subsequently, we demonstrate that {\bmc} can be easily adapted to GPU based LLM inference engines. Finally, we present a comprehensive ablation study, varying different hyper parameters and system knobs. 
 
\subsection{\textbf{Validation of Analytical Model}} 

Section \ref{sec:bmc} establishes that the optimal number of allocations for {\bmc} is given by:
$$T = \sqrt{N{\cdot}\frac{\alpha{\cdot}BW}{2{\cdot}\beta{\cdot}C}}~=~\sqrt{C^\prime{\cdot}N}$$
Empirically we measure the achieved bandwidth ($\alpha{\cdot}BW$)
for data copying and compute performance ($\beta{\cdot}C$) on the AMD Genoa server, and compute $C^\prime$ as 0.1. Using this the above equation for $T$, we get $T=\sqrt{0.1*N}$. 
Figure~\ref{OPT_MODEL_STUDY} shows  the normalized latency (latency of attention block in a single layer and for a single batch of requests), as the number of allocations is varied, but keeping a fixed context length of $N=512$ and a batch size of $8$. The lowest latency occurs at $T=8$  across all OPT models.  This matches with $T~\approx~\sqrt{0.1*512}$.  Other context lengths and batch sizes also behave in a similar way and their results are not shown here due to space limitations.

\begin{figure}[htb]
\centering
    \includegraphics[height=3cm]{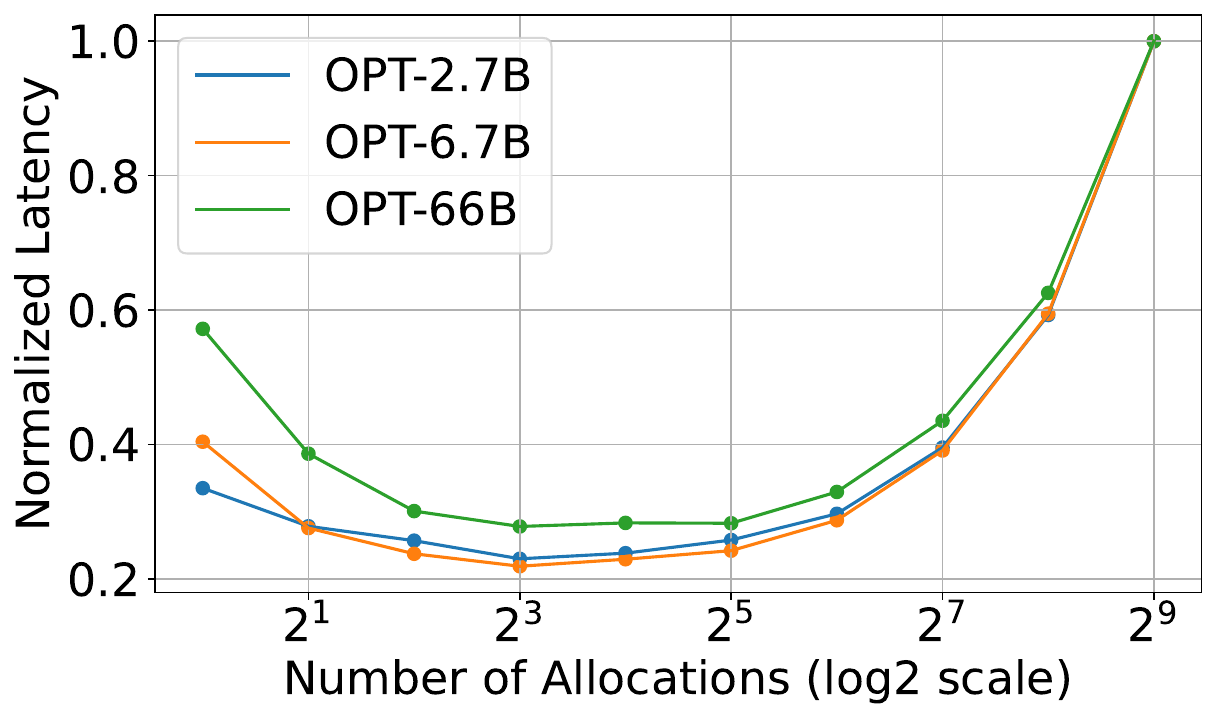}
 \caption{Attention Block Latency of OPT Models ($B=8$, $N=512$).}
        \label{OPT_MODEL_STUDY}
    \end{figure}

\begin{figure}[htb]
\centering
	\includegraphics[width=8.25cm]{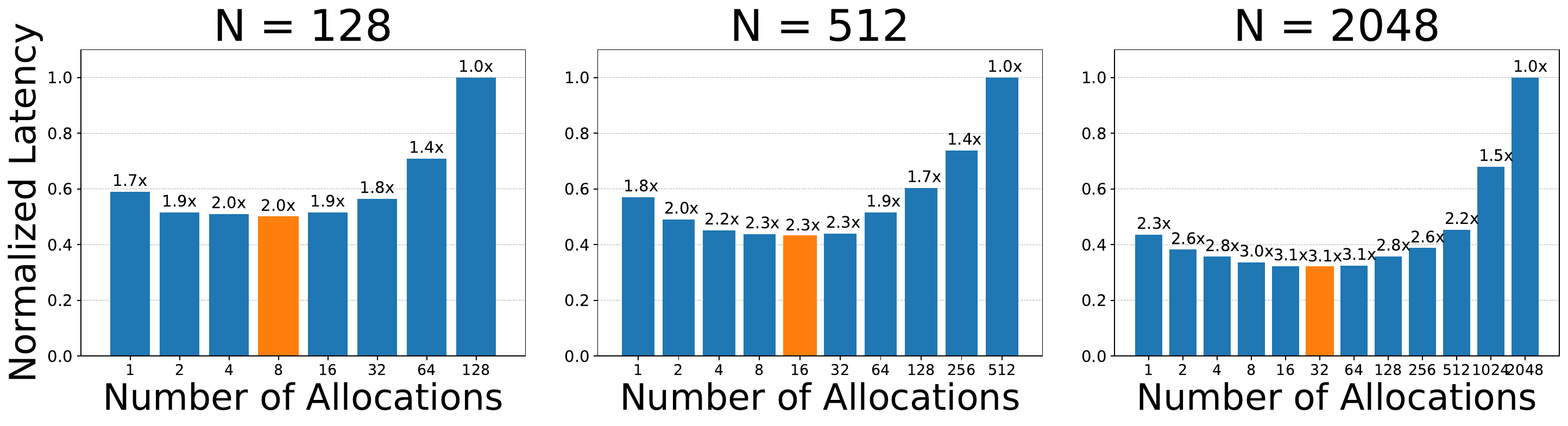}
	\caption{Attention Block Latency of OPT 6.7B ($B=8$). }
        \label{LLAMA_MODEL_NTOKENS}
    \end{figure}
    
\ignore{
\begin{figure}[htb]
\centering
	\includegraphics[width=8cm, height=3.5cm]{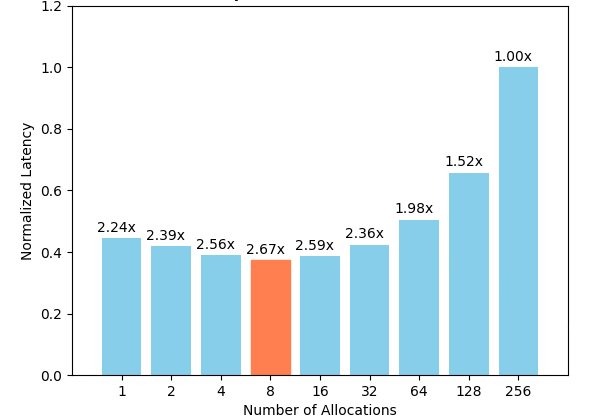}
	\caption{Attention Block Latency of Llama7B: $N=256$ and $B=8$.}
        \label{LLAM_MODEL_256}
    \end{figure}
    \FloatBarrier

\begin{figure}[htb]
\centering
	\includegraphics[width=8cm, height=3.5cm]{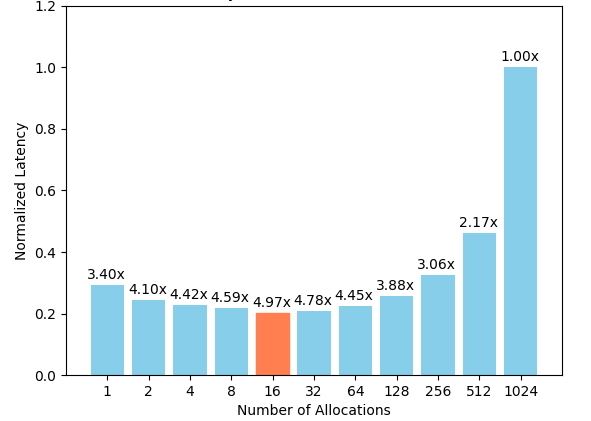}
	\caption{Attention Block Latency of Llama7B: $N=1024$ and $B=8$. }
        \label{LLAMA_MODEL_1024}
    \end{figure}
    \FloatBarrier

\begin{figure}[htb]
\centering
	\includegraphics[width=6cm, height=3.5cm]{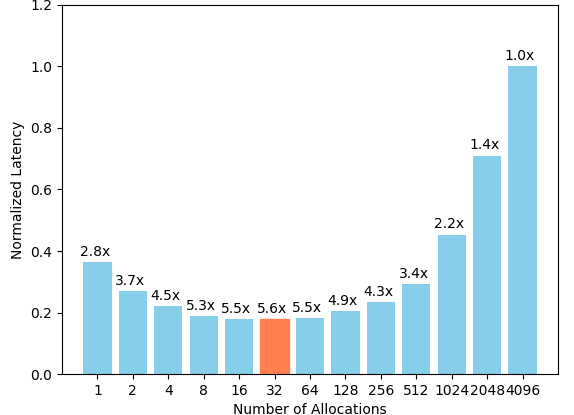}
	\caption{Attention Block Latency of Llama7B: $N=4096$ and $B=8$. }
        \label{LLAMA_MODEL_4096}
    \end{figure}
    \FloatBarrier
}

An interesting aspect of the optimal allocation strategy is that  $T$ is independent of the LLM  and the model parameters. This is validated across different OPT models in Figure~\ref{OPT_MODEL_STUDY}. 
\rnew{Next, we demonstrate how the best-performing $T$ changes with context lengths ($N=$ $128, 512$ and $2048$) 
for the OPT-6.7B model }(Attention Heads $H = 32$, Head Dimension $d = 128$). {Figure~\ref{LLAMA_MODEL_NTOKENS} presents} the normalized latency of the attention block computation using the {\bmc} approach as the number of allocations is varied. The number on top of each bar indicates the speedup (in attention block latency) compared to the baseline (iterative allocation). We observe that the lowest latency occurs when $T=8$, $T=16$ and $T=32$, respectively, when the context length ($N$) \rnew{ changes from 128, to 512, to 2048.} Specifically, when $N$ increases by a factor of 4, $T$ increases by a factor of 2, validating our analysis that $T\propto\sqrt{N}$. 

\ignore{ 
\begin{table}[htbp]
\centering
\begin{tabular}{|c|c|}
\hline
\textbf{Number of } & \textbf{Normalized } \\ 
\textbf{Allocations} & \textbf{Latency} \\ \hline
\hline
1 & 1.00 \\ \hline
2 & 0.87 \\ \hline
4 & 0.74 \\ \hline
8 & \textbf{0.72} \\ \hline
16 & 0.74 \\ \hline
128 & 1.28 \\ \hline
256 & 2.25 \\ \hline
512 & 4.02 \\ \hline
1024 & 7.11 \\ \hline
\end{tabular}
\caption{Attention Block Latency in Speculative Decoding.}
\label{tab:number_of_allocations_latency}
\end{table}
}

\begin{figure}[htbp]
    \centering
    \includegraphics[width=0.9\columnwidth]{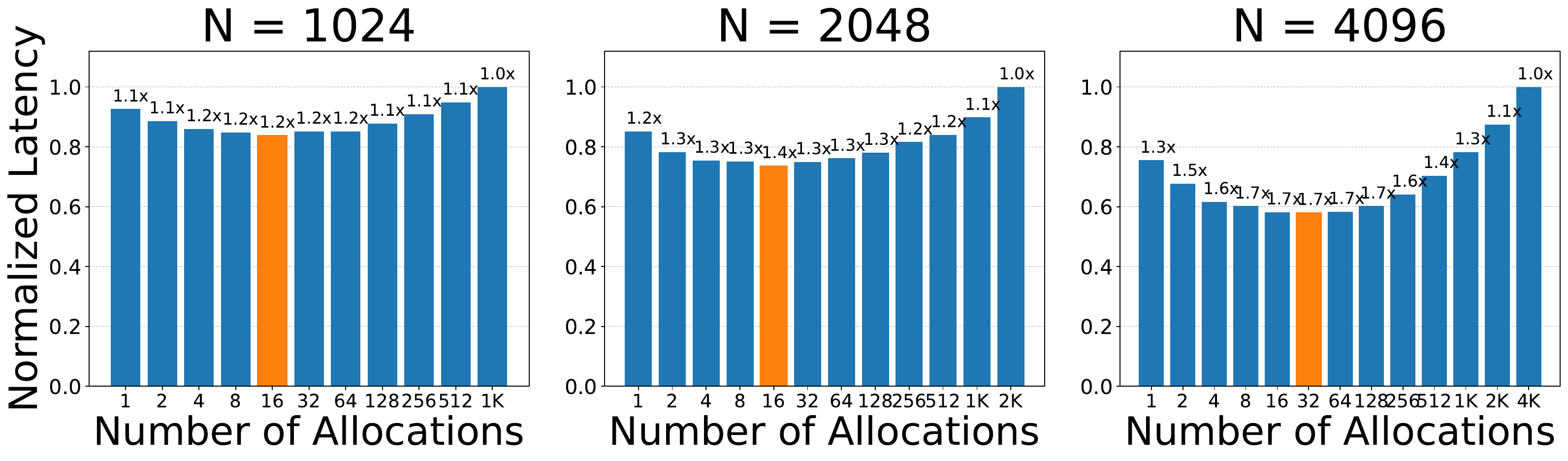}
    \caption{\textcolor{black}{Attention Block Latency for Llama3 8B, BS = 32 }}
    \label{fig:GQA-model-validation}
\end{figure}

\hpcarevise{
\subsubsection{Validation of Analytical Model for GQA}
Figure~\ref{fig:GQA-model-validation} shows the validation of our analytical model for GQA. The best performing $T$ values are 16 and 32, respectively, for $N=1024$ and $N=4096$.
}

\subsubsection{Validating Analytical Model for Speculative Decoding}
We consider the \todo{Llama 7B} with \rnew{32} heads, head size of 128, and maximum and a \todo {sequence length of 4096.}  The model speculates 26 candidates ($k=26$) and the number of accepted tokens ($m$) is 4. 
Table~\ref{tab:number_of_allocations_latency} reports the normalized attention block latency under SD, normalized w.r.t. the upfront allocation ($T=1$),  as the number of allocations is varied.  Our experiments demonstrate the existence of an optimal allocation between upfront and iterative allocation strategies,
"i.e.," the usefulness of our {\bmc} approach. 
We can observe that the minimum latency occurs for $T=8$.

\begin{table}[htbp]
\setlength{\tabcolsep}{4pt}
\centering
\caption{Attention Block Latency in Speculative Decoding}
\label{tab:number_of_allocations_latency}
\begin{small}
\begin{tabular}{|l|c|c|c|c|c|c|c|c|c|}
\hline
No. of       & 1 & 2 & 4 & 8 & 16 & 128 & 256 & 512 & 1024   \\ 
Allocs.  &   &  &  &  &  &  &  &  &  \\ \hline
Norm.   &  1.00 & 0.56 & 0.54 & \textbf{0.45} & 0.46 & 0.47 & 0.64 & 0.83 & 2.04 \\
Latency      &   &  &  &  &  &  &  &  &   \\ \hline
\end{tabular}
\end{small}
\end{table}

\subsection{\textbf{End-to-End Performance with Auto-Regressive Decoding}} \label{end-to-end-results}
In this section, we report the end-to-end speedup of {\bmc} for auto-regressive decoding.  
Speedup is calculated as the ratio of throughput (tokens/sec) achieved by {\bmc} (or other baselines) to that of the default iterative allocation.
Figure~\ref{OPT_MODEL_DIFFERENT_ALLOCS} compares {\bmc} with iterative and upfront allocation at a fixed batch size ($B=8$) and context length ($N=2048$). {\bmc} achieves up to $2.5\times$ speedup, with a geometric mean of $2.0\times$ over iterative allocation.

\ignore{
\noindent \textbf{Varying Context Length:} Figure~\ref{OPT_MODEL_FIXED_BS} shows the speedup of {\bmc} across OPT models with a fixed batch size ($B=8$) and context lengths ranging from 512 to 2048. Speedup generally increases with longer contexts, reaching up to $2.5\times$ across models.

\noindent \textbf{Varying Batch Size:} Figure~\ref{OPT_MODEL_DIFFERENT_BS} shows that with a fixed context length ($N=2048$) and batch sizes varying from 8 to 128, {\bmc} consistently outperforms the baseline allocation methods.
}

The performance of {\bmc} across different OPT models for varying context lengths and 
batch sizes (keeping the other parameter constant) are shown in  
\hpcarevise{Figure~\ref{OPT_MODEL_VARYING_CL_BS}.  Here the speedup is with respect to the Paged Attention (vLLM) model.
{\bmc} delivers speedup in the range of  1.20x~-- 1.47x over vLLM for different models, sequence lengths and batch sizes.} 
\ignore{
For extremely small models such as OPT125M the smaller dimensions and lower batch sizes lead to low memory operations making them less memory intensive.  On the other hand, extremely large models  have higher dimensions and the workload  tends to compute intensive, making the benefits with increased context length marginal. 
}

\begin{figure}[htb]
\begin{center}
	\includegraphics[height=2.5cm]{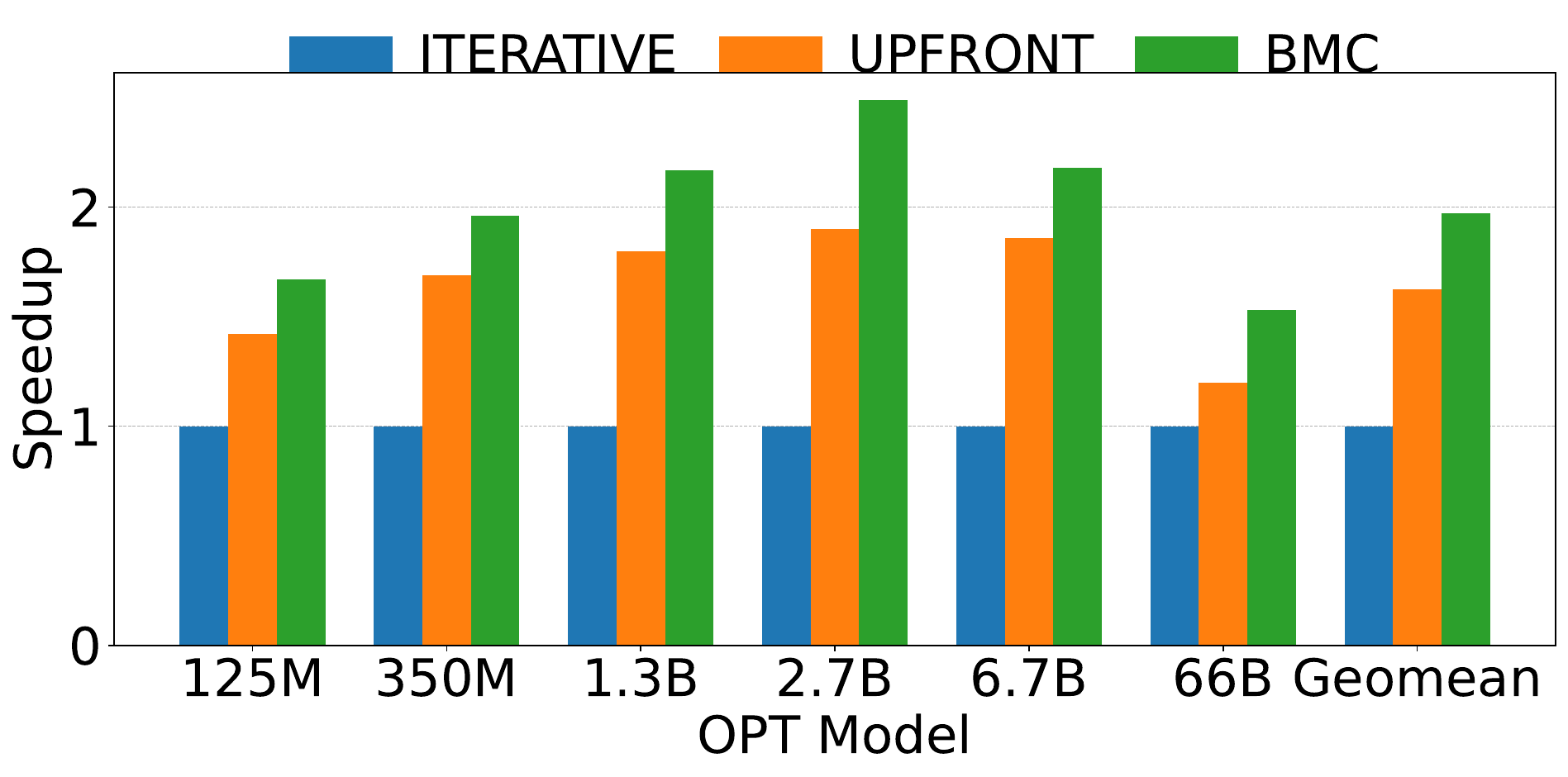}
        \caption{End-to-End Speedup Comparison of {\bmc} vs. Iterative and Upfront Allocation ($B=8$, $N=2048$).}
        \label{OPT_MODEL_DIFFERENT_ALLOCS}
    \end{center}
\end{figure}

\begin{figure}[htb]
\begin{center}
	\includegraphics[height=3.1cm]
    {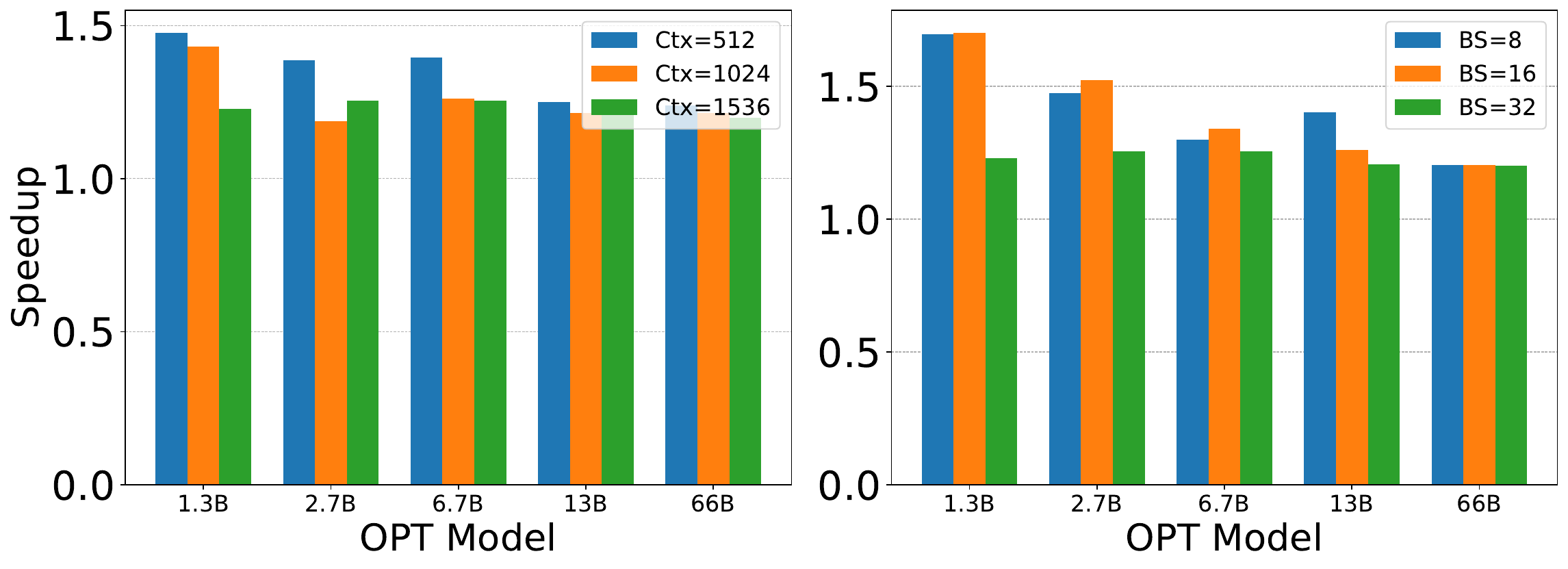}
            \caption{\textcolor{black}{End-to-End Speedup of BMC (over vLLM)  }}
        \label{OPT_MODEL_VARYING_CL_BS}
    \end{center}
\end{figure}

\hpcarevise{
\subsubsection{Performance with GQA } \label{sec:GQAPerf}
Table~\ref{tab:throughput-comparison} 
reports the performance of Hugging Face and {\bmc}, normalized to vLLM, on two GQA models (Llama-3 8B and Qwen-2 7B), for different use cases.
As can be seen from the table, {\bmc} achieves consistent performance gains over vLLM, in the range 1.12x~-- 1.68x in Llama 8B model and even higher gains (1.25x~--2.09x) for Qwen 8B model.
}

\begin{table}[ht]
  \centering
  \caption{\textcolor{black}{Throughput comparison for GQA models across representative use cases. Throughput normalized w.r.t. vLLM}}
  \label{tab:throughput-comparison}
  \footnotesize
  \setlength{\tabcolsep}{3.0pt}
  \renewcommand{\arraystretch}{1.05}
  \begin{tabular}{@{}l|ccc|SS|SS@{}}
    \toprule
   \multirow{2}{*}{Use Cases} 
  & \multicolumn{3}{c|}{Configuration} 
  & \multicolumn{2}{c|}{Llama 3 8B} 
  & \multicolumn{2}{c}{Qwen 2 7B} \\ 
    \cmidrule(lr){2-4}\cmidrule(lr){5-6}\cmidrule(lr){7-8}
    & Input & Output & Batch & {HF} & {BMC} & {HF} & {BMC} \\
    \midrule
                       & 128  & 128  & 1   & 0.99 & 1.59 & 1.49 & 1.78 \\
    Chat              & 128  & 128  & 32  & 0.82 & 1.34 & 1.28 & 1.54 \\
                      & 128  & 128  & 128 & 0.77 & 1.31 & 1.26 & 1.37 \\ \midrule
                      & 128  & 1920 & 1   & 0.41 & 1.41 & 0.59 & 1.29 \\
    Q\&A              & 128  & 1920 & 32  & 0.42 & 1.44 & 0.71 & 1.35 \\
                      & 128  & 1920 & 128 & 0.43 & 1.41 & 0.75 & 1.33 \\ \midrule
                      & 1920 & 128  & 1   & 0.45 & 1.63 & 0.91 & 1.95 \\
    Summarization     & 1920 & 128  & 32  & 0.50 & 1.68 & 1.05 & 2.03 \\
                      & 1920 & 128  & 128 & 0.48 & 1.65 & 1.01 & 1.96 \\\midrule
                      & 1920 & 1920 & 1   & 0.29 & 1.11 & 0.51 & 1.25 \\
    Long-form Gen.    & 1920 & 1920 & 32  & 0.31 & 1.15 & 0.55 & 1.29 \\
                      & 1920 & 1920 & 128 & 0.30 & 1.12 & 0.53 & 1.26 \\
    \midrule
    \multicolumn{4}{l|}{Geomean} & 0.66 & 1.35 & 0.82 & 1.59 \\
    \bottomrule
  \end{tabular}
\end{table}

\begin{table}[h] 
\setlength\tabcolsep{6pt}
\centering
\caption{\textcolor{black}{Speedup of {\bmc} (over Hugging Face) across models using  GSM8K dataset with $BS=32$.}}
\label{tab:attention_bmc_comparison_seq1024}
   \begin{small}
    \begin{tabular}{|c|c|c|c|c|c|}   \hline  
        Models  &  {OPT 1.3B} & {OPT 2.7B} & {OPT 6.7B} \\  
        \hline 
        {\bmc} Speedup & 3.02x & 3.46x & 3.57x  \\  \hline
    \end{tabular} 
    \end{small} 
\end{table}
\subsubsection{Highly Varying Sequence Lengths} \label{sec:VaryingSeqLength}
To evaluate the impact of variable sequence lengths, we use the GSM8K dataset ~\cite{cobbe2021gsm8k} and  report the end-to-end speedups of {\bmc} over Hugging Face. Across different OPT models, we achieve a speedup of 3.02x~--3.57x.

\subsection{\textbf{End-to-End Performance with Speculative Decoding}}\label{sec:SDperf}
\ignore{Speculative decoding is \textbf{not supported in the CPU implementation of  vLLM and DeepSpeed}. vLLM’s custom kernel implementation documentation explicitly lists limitations, available on their website, that prevent support for SD.  \revise{Furthermore, our manual verification confirms that vLLM's CPU backend for speculative decoding with Eagle ~\cite{li2024eagle} is not functional}. Hence, we }
We evaluate {\bmc} within the SpecBench framework~\cite{hemingkx_specbench}, a comprehensive benchmark and evaluation platform for speculative decoding (SD) built on HuggingFace. SpecBench includes a variety of tasks, including multi-turn conversations, question answering, and summarization. It supports a batch size of 1 and has a hardcoded token generation limit of 1024.

\begin{figure} [htb]
\centering
	\includegraphics[height = 3.2cm]{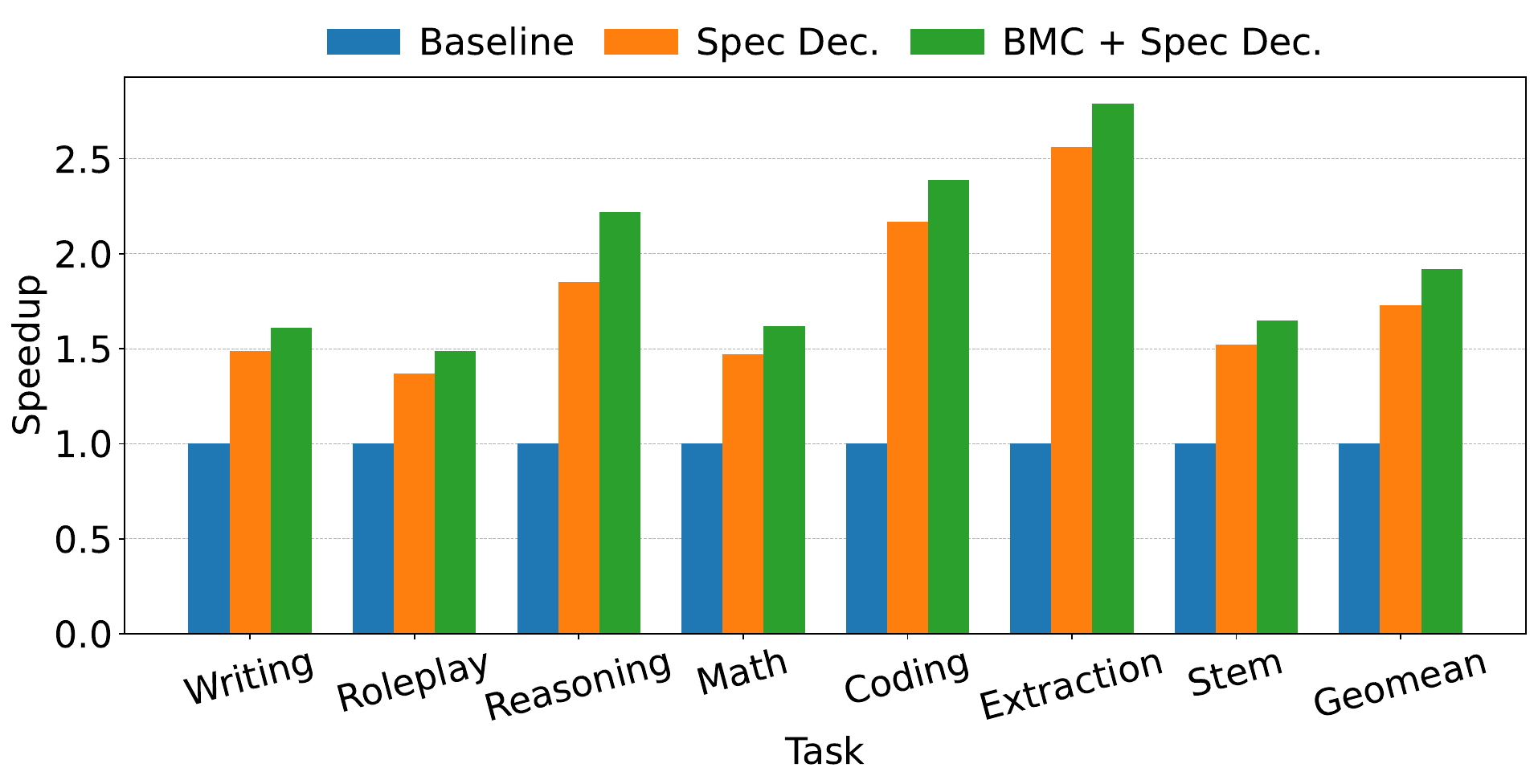}
        \caption{{\bmc} boosts speculative decoding performance across multi-task benchmarks (LLaMA2-7B, $B=1$).}
        \label{MultiTaskBenchmarks}
    \end{figure}

We use LLaMA2-7B as the base model, paired with draft model weights from Eagle~\cite{li2024eagle} using a candidate tree size of $k=26$. Figure~\ref{MultiTaskBenchmarks} shows performance across seven benchmark tasks in SpecBench.
As expected, SD provides a notable overall speedup of $1.73\times$ over standard auto-regressive decoding. Integrating {\bmc} into the SD pipeline yields a further performance gain, achieving an overall speedup of $1.90\times$. Despite SpecBench's batch size constraint, {\bmc} delivers an additional \todo{$1.1\times$} improvement, demonstrating its compatibility and benefits within SD frameworks.

\noindent \textbf{{\bmc} Performance over Parallel Draft Model (PARD):} 
Here we compare {\bmc} with a more recent PARD (PARallel Draft Model) framework~\cite{an2024pard} that minimizes draft model overheads with parallel drafting and achieves improved token acceptance \ignore{(5-7 tokens)}. 
\ignore {Table ~\ref{tab:speedup_comparison} reports the performance uplift of PARD and PARD+BMC over baseline.The performance benefits of {\bmc} over PARD is \textbf{1.28x~-- 1.35x}. }
\ignore {
The draft model used in SpecBench is 
sequential. Here we compare {\bmc} with a more recent PARD (PARallel Draft Model) framework~\cite{an2024pard} that minimizes draft model overheads with parallel drafting and achieves improved token acceptance \ignore {(5-7 tokens)}.} 
\revise{Figure~\ref{fig:SD_eval} demonstrates the performance improvements of BMC with PARD across using \texttt{humaneval} dataset~\cite{chen2021codex} across different models.  {\bmc} achieves a speedup of 1.23x~-- 1.39x}{Table ~\ref{tab:speedup_comparison} reports the performance uplift of PARD and PARD+BMC over baseline. The performance benefits of {\bmc} over PARD is \textbf{1.28x~-- 1.31x}. 

\begin{table}[h]  
\centering  
\begin{tabular}{|c|c|c|}  \hline
\textbf{Sequence} & \multicolumn{2}{c|}{\textbf{Speedup of Over Baseline}} \\ \cline{2-3}
\textbf{Length} & \textbf{PARD ~~~~} &  \textbf{PARD + BMC }  \\
\hline   
1024  & 5.44$\times$ & 7.18$\times$ \\   
2048  & 5.53$\times$ & 7.08$\times$ \\ \hline
\end{tabular}  
\caption{
Performance Comparison with PARD.}
\label{tab:speedup_comparison}  
\end{table}
}

\ignore{
As an inference serving solution, BMC\_MI launches improves performance with efficient data parallel instances achieving up to 11.11X performance over Auto-regressive mode and 2.52X over Speculative Decoding.

\begin{figure}[htb]
\centering
	
	\includegraphics[width=7cm]{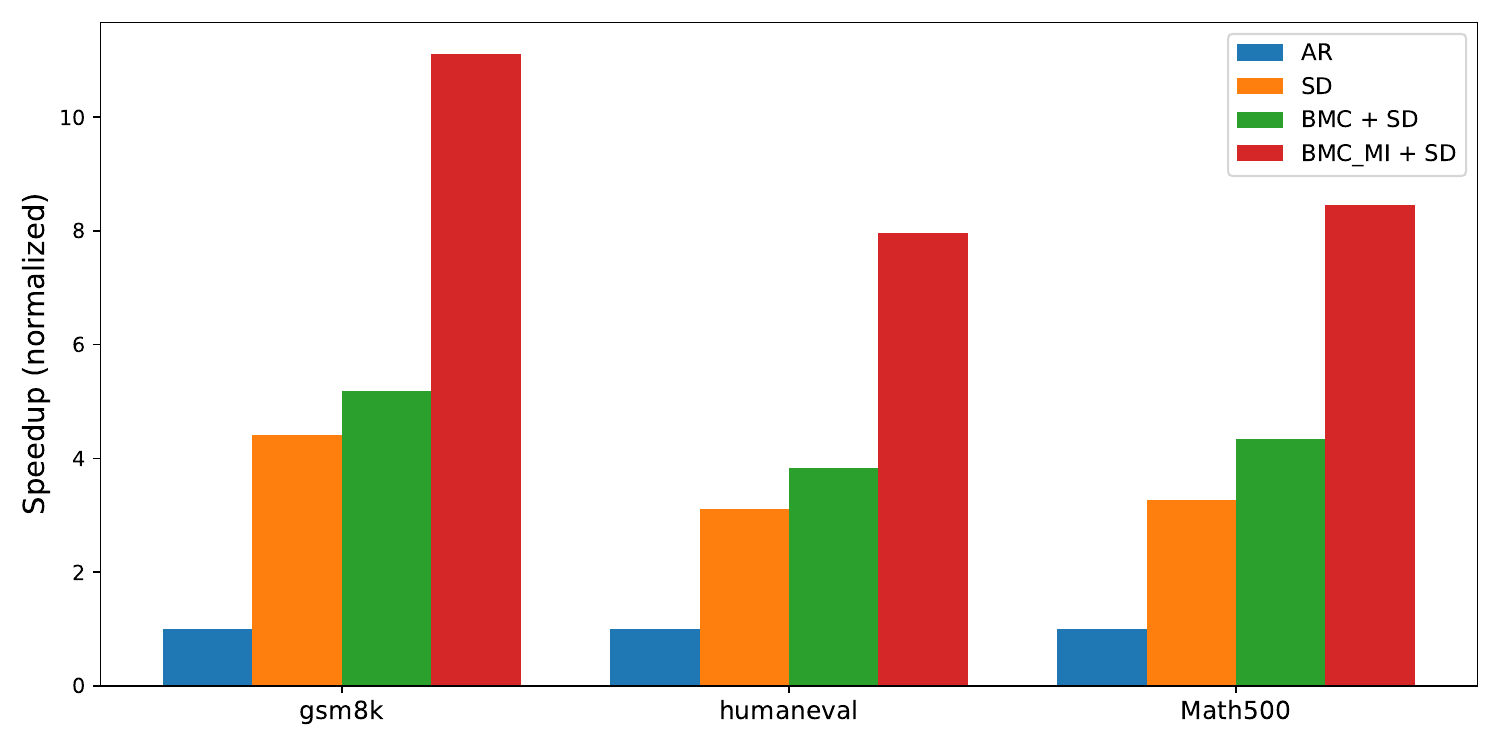}
	\caption{BMC boosts speculative decoding performance across benchmarks ($B=1$). }
        \label{fig:SD_MI}

    \end{figure}
}


\ignore{
\begin{figure}[htb]
\centering
	
	\includegraphics[width=7cm]{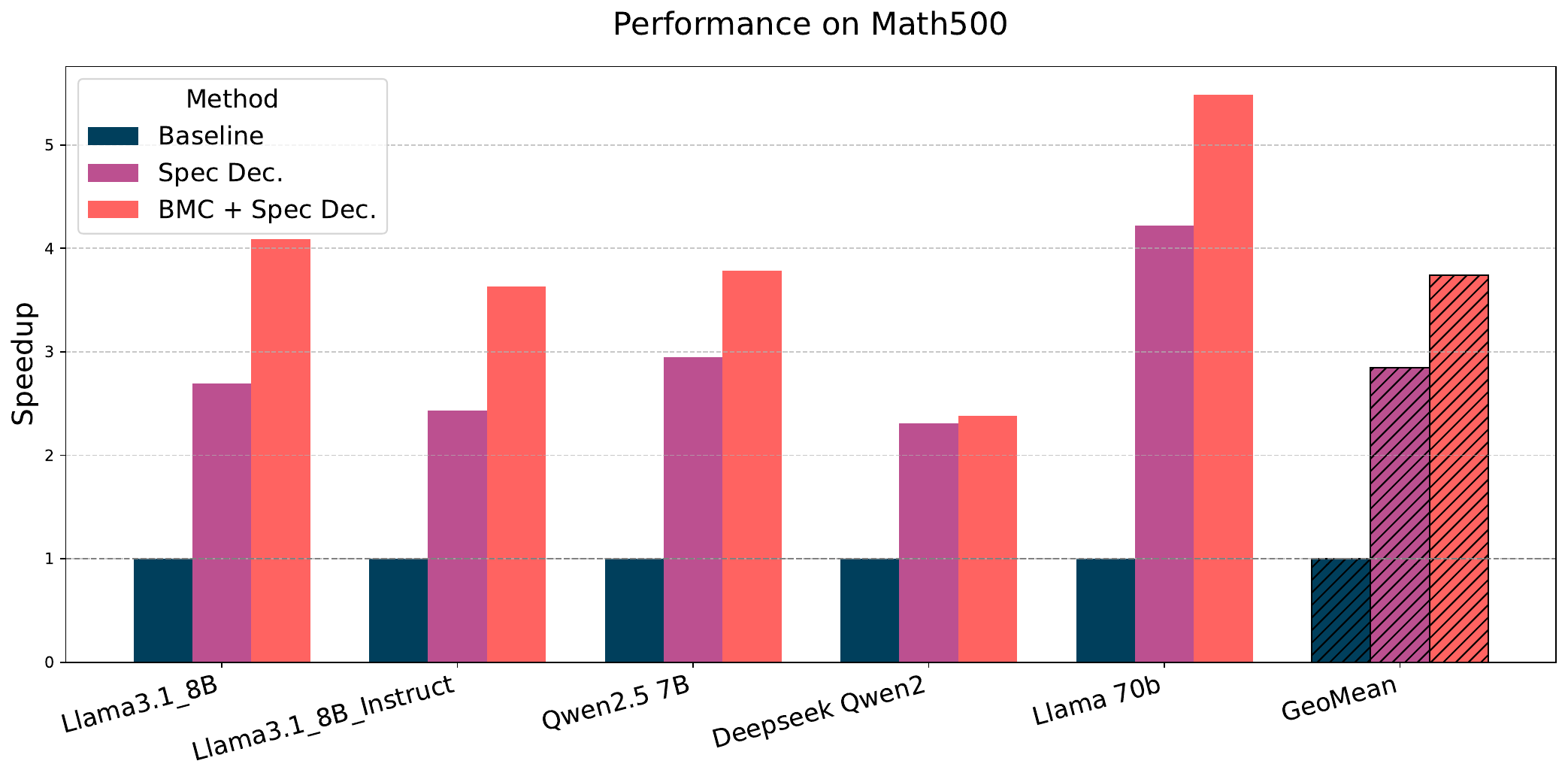}
	\caption{BMC Speedup with math500 ($B=1$). }
        \label{fig:SD_Math}

    \end{figure}
}

\begin{figure}[htb]
\centering
	\includegraphics[height=3.0cm,width=6.5cm]{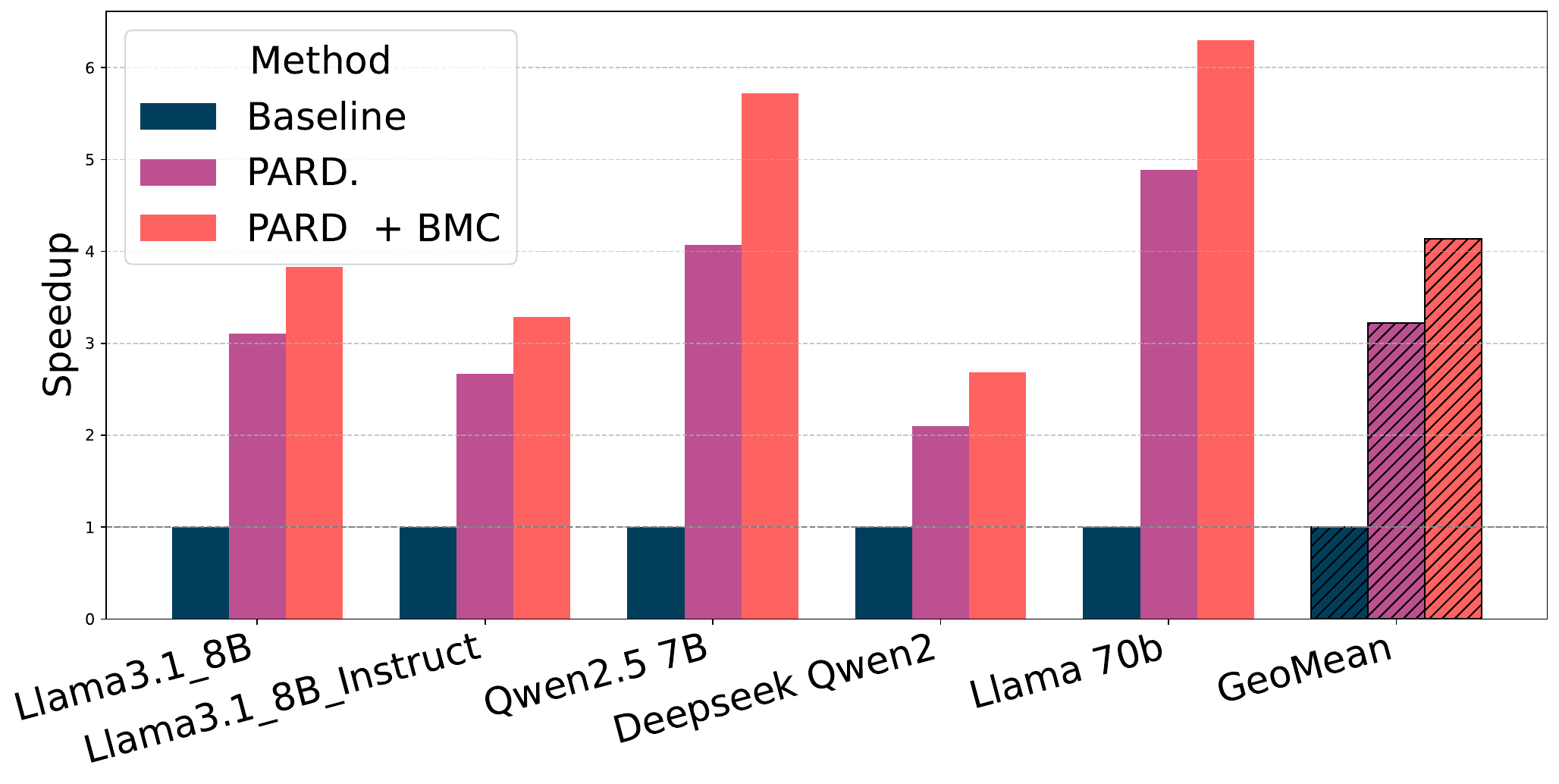}
	\caption{BMC  Speedup with humaneval ($B=1$). }
        \label{fig:SD_eval}
    \end{figure}

\subsection{\textbf{Comparison with SOTA Inference Servers Without Speculative Decoding}}\label{comp-vllm-deepspeed}
Here we compare end-to-end performance  of {\bmc} against the state-of-the-art inference servers\footnote{Another popular model used for CPU inference is \texttt{llama.cpp}.   Our evaluation on AMD EPYC Genoa (Zen4 architecture) CPUs, utilizing various backends for llama.cpp (OneAPI, Native, AMD BLIS) showing that llama.cpp is atleast \textbf{2.1x slower 
compared to BMC} across Llama 2 (7B/13B). 
As OPT models are not supported ~\cite{varunlmxd:LlamaCpp:OPT:2024} by llama.cpp, we did not include it in our quantitative comparison in Figure~\ref{E2E_Comparison}.}, namely 
DeepSpeed ~\cite{2022deepspeed} and vLLM (vllm)~\cite{Paged_Attention}. 
The Paged Attention ~\cite{Paged_Attention} scheme within the vLLM server was proposed for GPUs to specifically handle the memory capacity issues, and has also been adapted for CPUs, with a supporting implementation recently made available ~\cite{CpuPagedattention}. We use the recommended settings from ~\cite{vllm_cpu_installation} and the benchmark latency scripts ~\cite{vllm_benchmarks} for measuring the performance of vLLM. 

DeepSpeed is an open source deep learning optimization library~\cite{2022deepspeed} which 
achieves higher inference throughput by running multiple batches together and can support multi-instance deployments. 
To make a fair comparison, we also adapted similar optimizations to {\bmc}.
Further, we report the performance of a multi-instance version of {\bmc},  referred to as \bmc\_MI. 

\ignore {Another popular model used for CPU inference is \texttt{llama.cpp}.  However, findings from ~\cite{LlamaCppDiscussion13849_VishalZetta} on a 64-core Zen4 system indicate that vLLM significantly outperforms llama.cpp. Our own evaluation on AMD EPYC Genoa (Zen4 architecture) CPUs, utilizing various backends for llama.cpp (OneAPI, Native, AMD BLIS), aligns with this observation, showing that llama.cpp is  \textbf{1.59x~-- 1.81x slower 
compared to vLLM} across a range of models, including Llama 2 (7B/13B), Qwen-7B and Phi-2.  As OPT models are not supported ~\cite{varunlmxd:LlamaCpp:OPT:2024} by llama.cpp, these results are not shown in Figure~\ref{E2E_Comparison}.} 

    \begin{figure}[htb]
    \begin{center}
        \includegraphics[height=3.25cm,width=8cm]
        {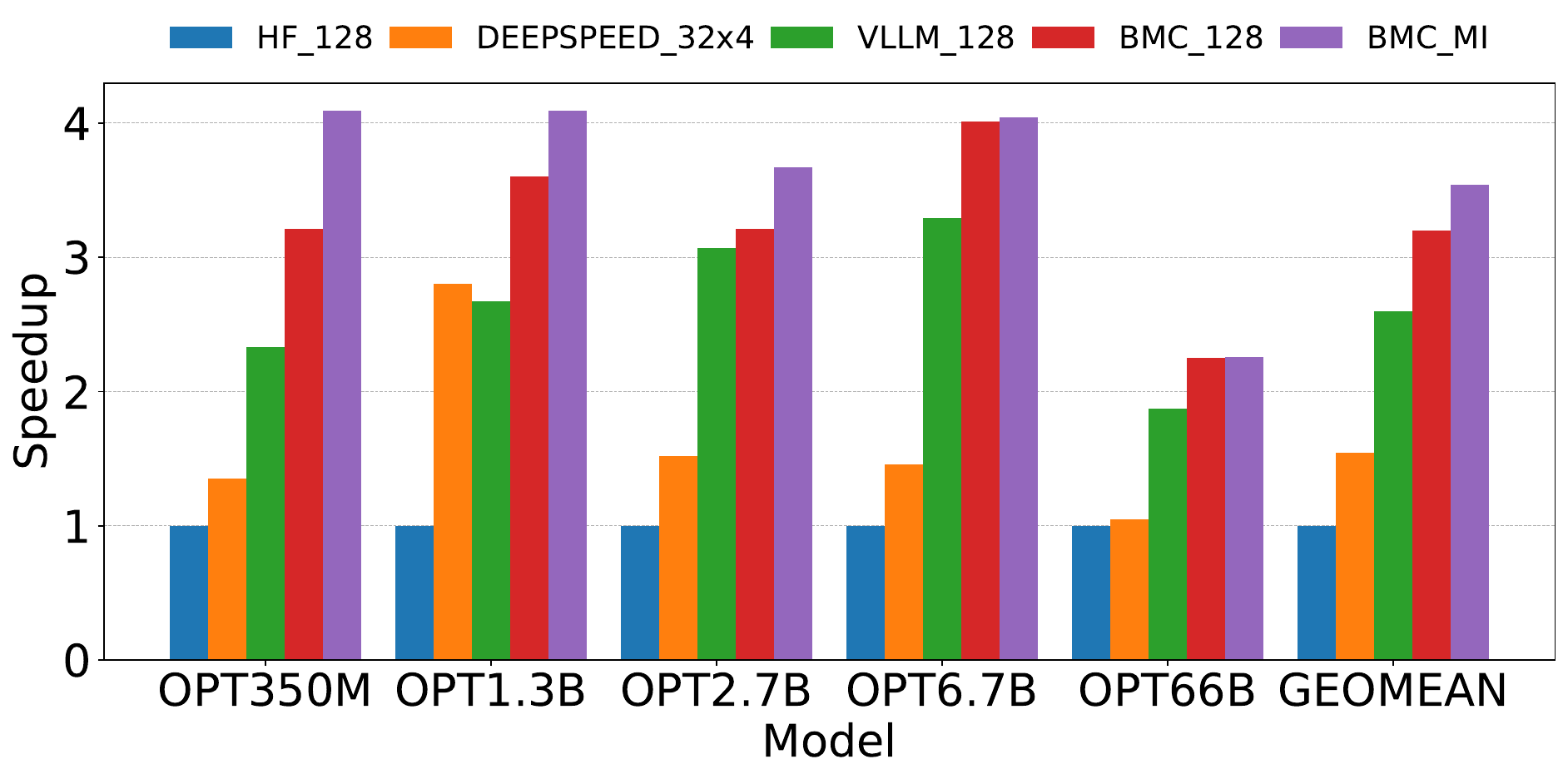}
        \caption{\textcolor{black}{End-to-End Performance Comparison of HF, DeepSpeed, vLLM, BMC and BMC\_MI ($B=128$, $N=2048$).}}
            \label{E2E_Comparison}
    \end{center}
    \end{figure}


Figure~\ref{E2E_Comparison} reports the end-to-end speedup of different inference methods, normalized to the throughput of the HuggingFace baseline. Both {\bmc} and {\bmc\_MI} consistently outperform HuggingFace, vLLM, and DeepSpeed across all models. The geometric mean speedups of {\bmc} are
\new{$3.2\times$, $2.07\times$, and $1.23\times$}{} over HuggingFace, DeepSpeed, and vLLM, respectively. {\bmc\_MI} achieves corresponding speedups of \new{$3.54\times$, $2.29\times$, and $1.36\times$.}{}

\subsection{\textbf{Reasons for {\bmc}'s Performance}}\label{sec:BMCAnalysis}
In this section we analyze the reasons for BMC performance improvements. 
Table~\ref{tab:perf_by_subsystem_opt2.7b} 
reports the break-up of attention block latency in terms of the time for KV cache copy, memory allocation for $K$ and $V$ tensors, and the SDPA time.   All times are normalized w.r.t. block attention time for the iterative scheme. The  memory allocation time decreases significantly for upfront allocation  and {\bmc} scheme, by $1536\times$ and $219\times$, respectively.  While upfront allocation does not incur any copy overhead, {\bmc}'s copy time is atleast $204\times$ lower than that of the iterative approach. These benefits, in turn, result in the improved performance of {\bmc} over the baseline.
\revise{While vLLM incurs very little memory allocation overhead, the time for SDPA is higher, making the overall attention block latency higher than that for {\bmc}.}{}

\begin{table}[htbp]
  \centering
    \caption{ Break-up of Attention Block Latency} 
  \label{tab:perf_by_subsystem_opt2.7b}
\ignore{
\begin{tabular}{|l|l|r|r|r|}
    \hline
         Model & Scheme & Memory     & Cache   & SDPA \\
               &        & Allocation & Copying &       \\ \hline
            & Iterative & 1.0000   & 0.488 & 0.682 \\ \cline{2-5}
OPT 2.7B    & Upfront   & 0.00059  & 0.00000 & 1.22 \\ \cline{2-5}
            & {\bmc}   & 0.0046 & 0.0023 & 0.6345 \\ \cline{2-5}
            & {vLLM}   & 0.0053 & 0.0  & 0.706 \\ \hline
    \end{tabular}%
    }
  \begin{tabular}{|l|l|r|r|r|r|} \hline
Model & Scheme & Memory     & Cache   & SDPA & Total \\  
      &        & Allocation & Copying &     &   \\ \hline
    & Iterative &	0.4608 &	0.2249 &	0.3143 &	1.0000 \\ \cline{2-6}
OPT 2.7B    & Upfront   &	0.0003 &	0.0 &	0.5622 &	0.5625 \\ \cline{2-6}
    & vLLM   &	0.0024 &	0.0 & 	0.3253 &	0.3278 \\ \cline{2-6}
    & {\bmc}   &	0.0021 &	0.0011 &	0.2924 &	\textbf{0.2956} \\ \hline
\end{tabular}
\end{table}%

It is interesting to note that the SDPA time of our {\bmc} is in fact \emph{lower} (by more than 3\%) than that of the iterative approach, even though the {\bmc} approach is doing more computation (due to the zero-padded rows).  Further profiling  of the execution time revealed that the BMM in SDPA execution time involves a Just-In-Time (JIT) compilation~\cite{jit}  that  specializes the BLAS call for the given matrix dimension.  While the iterative approach incurs this JIT compilation overhead in every iteration,  it is limited to $T$ iterations for {\bmc}, the iterations in which the memory allocation for $K$ and $V$ is performed.

Table~\ref{tab:cache_tlb_misses_opt1.3b}
reports the number of page faults and TLB misses incurred by different schemes. We experimented with the normal page size (4KB pages) and large pages (2MB pages). 
The number of (minor) page faults decrease by nearly two orders of magnitude with both upfront allocation and {\bmc} scheme. This trend holds even when large pages are enabled.  We observe that the  {\bmc} scheme reduces the number of TLB misses by nearly 50\%.  
These observations align well with the break-up of attention computation time reported earlier.  
\begin{table}[htbp]
  \centering
  \caption{Page Faults and TLB Misses for \rnew{OPT~6.7B} Model.}
\label{tab:cache_tlb_misses_opt1.3b}
\begin{tabular}{|p{0.8cm}|l|l|r|r|r|}\hline
Page          & Scheme & \multicolumn{2}{c|}{Page Faults} & TLB misses   \\ \cline{3-4}
Size              &           &  Major    &  Minor       &         \\ \hline 
Small   & Iterative &      261     & 2,190,549,334  & 13,826,312,862 \\ 
Pages   & Upfront   &      116     &  25,944,681   & 8,961,418,388  \\ 
(4KB)   &  BMC      &      159     &  82,402,477   &  6,950,879,993 \\ \hline
Large   & Iterative &      208     &  38,668,153  &    572,561,285 \\ 
Pages   & Upfront   &      220     &    1,066,370    &   332,346,554 \\ 
(2MB)   & BMC       &        160  &   2,145,723    &  277,031,460   \\ \hline
\end{tabular}
\end{table}



\revise{
Are {\bmc}'s  performance gains  primarily a result of addressing the limited memory bandwidth of CPU systems? We present the performance gains  when only 8 (out of 2~$\times$~96) cores of the experimental 
platform is used. With no other application running, the 8 cores enjoy the full memory  memory bandwidth of 460.8 GBps.  Even with this sufficiently large memory bandwidth, the performance of {\bmc} is 1.25x to 2.25x  better (see Table~\ref{tab:vllm_bmc_comparison}), demonstrating that the benefits of {\bmc} are not necessarily coming from a bandwidth-limited system on which LLM is implemented.}{}

\begin{table}[h]  
\setlength\tabcolsep{3.0pt}
    \centering  
    \caption{{\bmc} Improvement (over vLLM) when using 8 cores and full system memory bandwidth.}
     \label{tab:vllm_bmc_comparison} 
    \begin{small}
    \begin{tabular}{|c|c|c|c|c|c|}   \hline  
        Batch Size & {OPT 350M} & {OPT 1.3B} & {OPT 2.7B} & {OPT 6.7B} \\  
        \hline  
        BS=8  &  1.55x & 2.25x & 1.50x & 1.79x \\ \hline  
        BS=32 &  1.56x & 1.60x & 1.72x & 1.25x \\  \hline
    \end{tabular} 
    \end{small} 
\end{table}

\ignore{
\begin{table}[htbp]
  \centering
\begin{tabular}{|p{0.8cm}|l|l|r|r|r|}\hline
Page          & Scheme & \multicolumn{2}{c|}{Page Faults} & TLB misses   \\ \cline{3-4}
Size              &           &  Major    &  Minor       &         \\ \hline \hline
Small   & Iterative &      341     & 676,773,780  & 5,611,138,736 \\ 
Pages   & Upfront   &      291     &  1,506,240   & 3,126,215,897 \\ 
(4KB)   &  BMC      &      129     &  8,176,719   & 2,300,389,421 \\ \hline
Large   & Iterative &      76     & 27,399,991   &   604,269,011 \\ 
Pages   & Upfront   &      429     &    152,719   &   235,957,089 \\ 
(2MB)   & BMC       &      209     &    693,918   &   245,659,782 \\ \hline
\end{tabular}
\label{tab:cache_tlb_misses_opt1.3b}
\end{table}
}

\hpcarevise{
\subsection{\textbf{{\bmc}'s Attention Performance}}\label{sec:SDPA Analysis}
We  include speedups in attention kernel execution time for {\bmc}'s contiguous allocation (over vLLM's non-contiguous allocation) for different OPT models (Table~\ref{tab:attention_bmc_comparison_seq1024}). {\bmc}  achieves speedups in the range of,  1.07x~-- 1.57x over custom non-contiguous paged attention kernels.  With BMC Multi-Instance (\emph{BMC\_MI}) the speedups are even higher (1.59x~-- 2.64x), as there is a balance in the per core and system utilization achieved by the OneDNN library.  
}

\begin{table}[h]
\setlength\tabcolsep{6pt}
\centering
\caption{Speedup in attention latency of contiguous allocation (BMC and BMC\_MI) over non-contiguous allocation (vLLM) across OPT models (batch size = 32, sequence length = 1024).}
\label{tab:attention_bmc_comparison_seq1024}
\begin{small}
\begin{tabular}{|c|c|c|c|c|c|} \hline
OPT Model & 350M & 1.3B & 2.7B & 6.7B & 13B \\ \hline
BMC       & 1.31x & 1.27x & 1.07x & 1.57x & 1.30x \\ \hline
BMC\_MI   & 1.75x & 2.68x & 1.27x & 2.47x & 1.59x \\ \hline
\end{tabular}
\end{small}
\end{table}

\subsection{\textbf{GPU Results}}\label{sec:GPU-results}

While {\bmc} was  evaluated on CPUs till now, it can also be integrated with LLM inference engines on GPUs.  In this section, we present additional experimental results on MI210 GPUs (with 128 GB RAM and ROCm 6.3). 

\revise{First we demonstrate the efficacy of our analytical model. Figure~\ref{fig:analyticalmodel_gpu} presents the attention block latency of OPT 13B model with Batch Size=32. We observe that the best-performing configuration occurs at $T=\sqrt{N}$. Similarly for speculative decoding  with candidate size=26 and acceptance length=8, the attention block latency of OPT 13B model for a Batch Size=32 and Sequence length of 4096 is shown in \ref{tab:SpecDec_latency}. \ignore{We also  evaluate the KV Cache Update Performance improvement of OPT 13B model, across different batch sizes but with Fixed Token budget of Cache Size 1024. BMC repurposes redundant compute to eliminate compaction for every drop, there by improving the KV update speedup as shown in ~\ref{tab:KV_speedup}}}.

\begin{figure}[htb]
\centering
	\includegraphics[width=8.25cm]{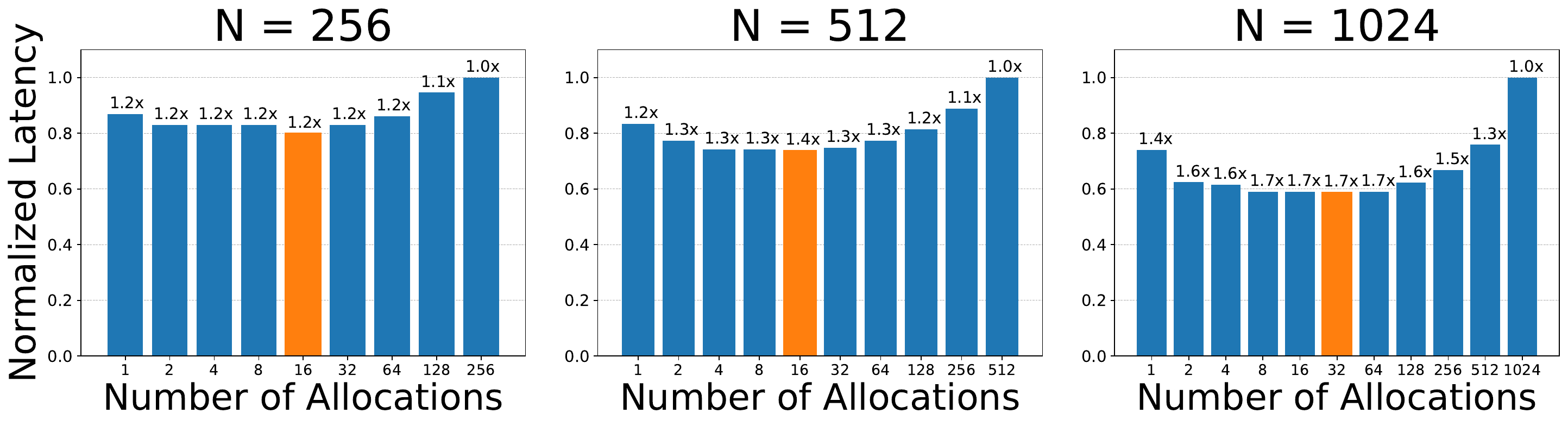}
	\caption{Attention Block Latency of OPT 13B ($B=32$). }
        \label{fig:analyticalmodel_gpu}
    \end{figure}

\begin{table}[htbp]
\setlength{\tabcolsep}{3pt}
\centering
\caption{Attention Block Latency in Spec. Decoding.}
\label{tab:SpecDec_latency}
\begin{small}
\begin{tabular}{|l|c|c|c|c|c|c|c|c|c|}
\hline
No. of       & 1 & 2 & 4 & 8 & 16 & 32 & 64 & 128 & 256   \\ 
Allocs.  &   &  &  &  &  &  &  &  &  \\ \hline
Norm.   &  1.00 & 0.65 & 0.55 & 0.509 & \textbf{0.501} & 0.503 & 0.52 & 0.57 & 0.74 \\
Latency      &   &  &  &  &  &  &  &  &   \\ \hline
\end{tabular}
\end{small}
\end{table}

\ignore{
\begin{table}[htbp]
\setlength{\tabcolsep}{4pt}
\centering
\begin{small}
\begin{tabular}{|l|c|c|c|c|c|c|c|c|c|}
\hline
Batch size       & 1 & 2 & 4 & 8 & 16 & 32 & 64 & 128    \\ 
  &   &  &  &  &  &  &  &  \\ \hline
Speed.   &  1.35 & 1.39 & 2.34 & 4.06 & 7.49 & 14.37 & 21.57 & 50.85 \\
Up      &   &  &  &  &  &  &  &    \\ \hline
\end{tabular}
\end{small}
\caption{KV Update Speedup ( Random Evictions for Cache Size 1024) with Token Dropping.}
\label{tab:KV_speedup}
\end{table}
}

\revise{Finally we present end-to-end performance of {\bmc} for batch size  32, 64 and 128 and with sequence length, respectively,  1024, 512 and 256.  The speedup results of {\bmc} on GPUs, obtained by comparing the optimized implementations to the baseline, are shown in Table~\ref{BMC-on-GPU}.  The speedup for large models ranges from 1.41$\times$  to 1.98$\times$. }{}
\ignore{
Our evaluation \todo{for end to end inference } utilizes batch sizes of 32 and 64 with sequence lengths of 1024 and 512, respectively.The speedup results of {\bmc} on GPUs, obtained by comparing the optimized implementations to the baseline, are shown in Table~\ref{BMC-on-GPU}.  The speedup ranges from 1.41$\times$ for large models to 3.95$\times$ (for OPT-125M with batch size 128). }
%

\begin{table}[hbt]
\begin{center}  
\caption{End-to-End Performance of {\bmc} on GPUs.}
\label{BMC-on-GPU}
\begin{tabular}{|l|c|c|c|} \hline
Model  & \multicolumn{3}{c|}{Speedup} \\ \cline{2-4}
 & $N = 256$  
 & $N = 512$ 
 & $N = 1024$   \\ 
 &  $B = 128$ 
 &  $B = 64$ 
 & $B = 32$  \\ \hline 
OPT 350M     & 1.98$\times$    & 1.61$\times$  & 1.62$\times$ \\  
OPT 1.3B     & 1.90$\times$    & 1.55$\times$  & 1.35$\times$  \\  
OPT 2.7B     & 1.72$\times$    & 1.50$\times$  & 1.27$\times$  \\  
OPT 6.7B     & 1.76$\times$    & 1.69$\times$  & 1.41$\times$  \\ \hline 
\ignore{
OPT 125M (125 million)    & 3.95    & 3.21  & 1.87 \\  
OPT 350M (350 million)    & 1.98    & 1.61  & 1.62 \\  
OPT 1.3B (1.3 billion)    & 1.90    & 1.55  & 1.35  \\  
OPT 2.7B (2.7 billion)    & 1.72    & 1.50  & 1.27  \\  
OPT 6.7B (6.7 billion)    & 1.76    & 1.69  & 1.41  \\ \hline
}
\end{tabular}  
\end{center}  
\end{table}

\ignore{
The speedup results of {\bmc} on GPUs, obtained by comparing the optimized implementations to the baseline, are shown in Table~\ref{BMC-on-GPU}.  The speedup ranges from 1.41$\times$ for large models to 3.95$\times$ (for OPT-125M with batch size 128). }

\ignore{
\subsubsection{Speedup Results Over Baseline for Sequence Length = 512 and Batch Size = 64}
  
\begin{center}  
\begin{tabular}{@{}ll@{}}  
\toprule  
Model (Parameters) & Speedup \\ \midrule  
OPT 125M (125 million)    & 3.21        \\  
OPT 350M (350 million)    & 1.61        \\  
OPT 1.3B (1.3 billion)    & 1.55        \\  
OPT 2.7B (2.7 billion)    & 1.50        \\  
OPT 6.7B (6.7 billion)    & 1.69        \\ \bottomrule  
\end{tabular}  
\end{center}  

\subsubsection{Speedup Results Over Baseline for Sequence Length = 1024 and Batch Size = 32}
 
\begin{center}  
\begin{tabular}{@{}ll@{}}  
\toprule  
Model  & Speedup \\ \midrule  
OPT 125M (125 million)    & 1.87        \\  
OPT 350M (350 million)    & 1.62        \\  
OPT 1.3B (1.3 billion)    & 1.35        \\  
OPT 2.7B (2.7 billion)    & 1.27        \\  
OPT 6.7B (6.7 billion)    & 1.41        \\ \bottomrule 
\end{tabular}  
\end{center}  
}

\subsection{Ablation Studies} 

\subsubsection{Larger Sequences}
\rnew{We explore the benefits of {\bmc}  for large sequence length 
(up to 16,384 tokens) and a batch size of 4 and 16.
}
For this, we use a head size of 32, a head dimension of 16.
BMC achieves a speedup (in SDPA execution time) of 
$10.58\times$  and 
$19.67\times$, 
respectively for ($B=4$) and ($B=16$), over HuggingFace.
\new{{\bmc}'s performance gains increase with both batch size and sequence length. 
Additionally at sequence length 8192, {\bmc} achieves 1.2× end-to-end speedup (over vLLM) for Llama-3 8B model.}


\begin{table}[h]
\caption{\textcolor{black}{{\bmc} Performance with Larger Sequence Lengths.}}
\begin{center}
\begin{tabular}{|l|c|c|c|c|c|}
\hline
Seq. Length & 1024 & 2048 & 4096 & 8192 & 16384 \\ \hline
$B=4$  & 4.21$\times$  & 4.33$\times$  & 6.04$\times$  & 7.69$\times$  & 10.58$\times$ \\ 
$B=16$ & 4.38$\times$  & 3.92$\times$  & 8.48$\times$  & 10.85$\times$ & 19.67$\times$ \\ \hline
\end{tabular}
\label{tab:LargeSeq}
\end{center}
\end{table}

\subsubsection{Compatibility with different backends}

\ignore{ 
\begin{table}[hbt]
\setlength{\tabcolsep}{2pt}
\centering
\begin{small}
\begin{tabular}{|l|c|c|c|c|c|c|c|c|c|c|} \hline  
No. of &  1   &  2   &  4   &  8   &  16  &  32  &  64 &  128 &  256 & 512 \\
Allocs. &     &      &      &      &      &      &     &      &     &       \\ \hline
Eager   & 1.000 & 0.921 & 0.857 & 0.819  & 0.837 & 0.862 &  0.957 & 1.067 &  1.243 & 1.728 \\
Fused    & 1.000  & 1.057  & 0.964 & 0.851 & 0.846 & 0.907 & 0.917 & 1.032 & 1.044 & 1.193 \\ \hline
\ignore{ 
1 & 1.000 & 12.000 \\
2 & 0.921 & 1.057 \\
4 & 0.857 & 0.964 \\
8 & 0.819 & 0.851 \\
16 & 0.837 & 0.846 \\
32 & 0.862 & 0.907 \\
64 & 0.957 & 0.917 \\
128 & 1.067 & 1.032 \\
256 & 1.243 & 1.044 \\
512 & 1.728 & 1.193 \\ \hline 
}
\end{tabular}
\end{small}
\caption{BMC Perfomance with Different Backends.}
\label{tab:performance_comparison}
\end{table}
}

\revise{We compare how {\bmc} performs with different backends, specifically the attention block computation using eager mode (runs SDPA operations sequentially using standard libraries) and Custom Kernels (fuses the operations in SDPA as in FlashAttention~\cite{dao2022flashattention}).  The performance trend is similar, and both backends achieve the best performance when $T~\propto~\sqrt{N}$.}{
Table~\ref{tab:performance_comparison} reports the normalized time of attention block computation using eager mode (runs SDPA operations sequentially using standard libraries) and Custom Kernels (fuses the operations in SDPA as in FlashAttention~\cite{dao2022flashattention}).  The performance trend is similar with different backends.}
\ignore{
\begin{table}[hbt]
\setlength{\tabcolsep}{2pt}
\centering
\begin{small}
\begin{tabular}{|l|c|c|c|c|c|c|c|c|c|c|} \hline  
No. of &  1   &  2   &  4   &  8   &  16  &  32  &  64 &  128 &  256 & 512 \\
Allocs. &     &      &      &      &      &      &     &      &     &       \\ \hline
Eager   & 1.000 & 0.921 & 0.857 & 0.819  & 0.837 & 0.862 &  0.957 & 1.067 &  1.243 & 1.728 \\
Mode    &     &      &      &      &      &      &     &      &      &       \\ \hline
SPDA   & 12.000  & 1.057  & 0.964 & 0.851 & 0.846 & 0.907 & 0.917 & 1.032 & 1.044 & 1.193 \\ \hline
\ignore{ 
1 & 1.000 & 12.000 \\
2 & 0.921 & 1.057 \\
4 & 0.857 & 0.964 \\
8 & 0.819 & 0.851 \\
16 & 0.837 & 0.846 \\
32 & 0.862 & 0.907 \\
64 & 0.957 & 0.917 \\
128 & 1.067 & 1.032 \\
256 & 1.243 & 1.044 \\
512 & 1.728 & 1.193 \\ \hline 
}
\end{tabular}
\end{small}
\caption{BMC Perfomance with Different Backends.}
\label{tab:performance_comparison}
\end{table}
}

\subsubsection{{\bmc} across Different Platforms} 

\ignore{ 
Based on the experimental results, we observe that the proposed BMC approach achieved up to 3.95$\times$ speedup over the HuggingFace baseline for the OPT 125m model. The BMC technique shows effective scalability, as evidenced by the performance improvements across a wide range of model sizes, from OPT 125M to OPT 6.7B.}

We experimented with an AMD Milan Server (7763 processor with 64 cores and with 256 GB memory), an AMD $Ryzen^{TM}$ Client (7840 processor with 8 cores and with 16 GB memory) and a 60-core Intel Xeon Platinum 8490H. We used the OPT~350M model  with a batch size $B=8$.  
\revise{}{Figures~\ref{sub:Milan} and \ref{sub:Ryzen} present the normalized attention block latency, for different token lengths ($N$), and  the number of allocations ($T$) is varied from 1 to $N$.} While the absolute performance varies across  platforms, \revise{we}{the trend remains the same. We} observe that for both target systems the best-performing number of allocations happens at $T=16$ for $N=1024$ tokens, $T=8$ for $N=256$, and $T=4$ for $N=64$.  We observe that the value of $T$ decrease by a factor of 2 when the value of $N$ decreases by 4.   
\ignore{ 
\begin{figure}[htb]
\centering
\begin{minipage}{3.75cm}
\centering
	\includegraphics[width=3.7cm]{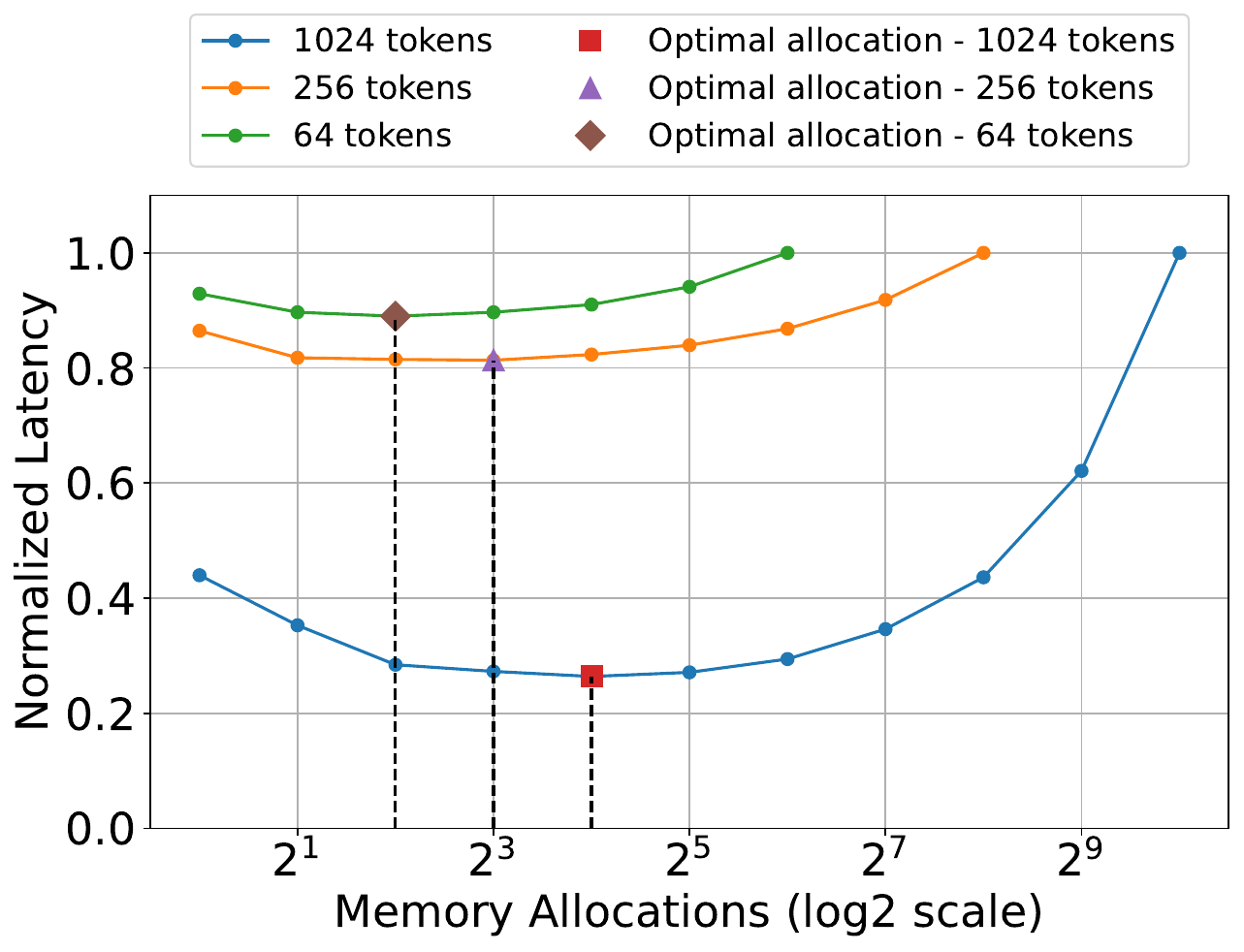}
    \captionof{figure}{Attention Block Latency on AMD Milan Server.}
        \label{sub:Milan}
\end{minipage}%
\hspace*{0.25cm}
\begin{minipage}{3.75cm}
 \centering
    \includegraphics[width=3.7cm]{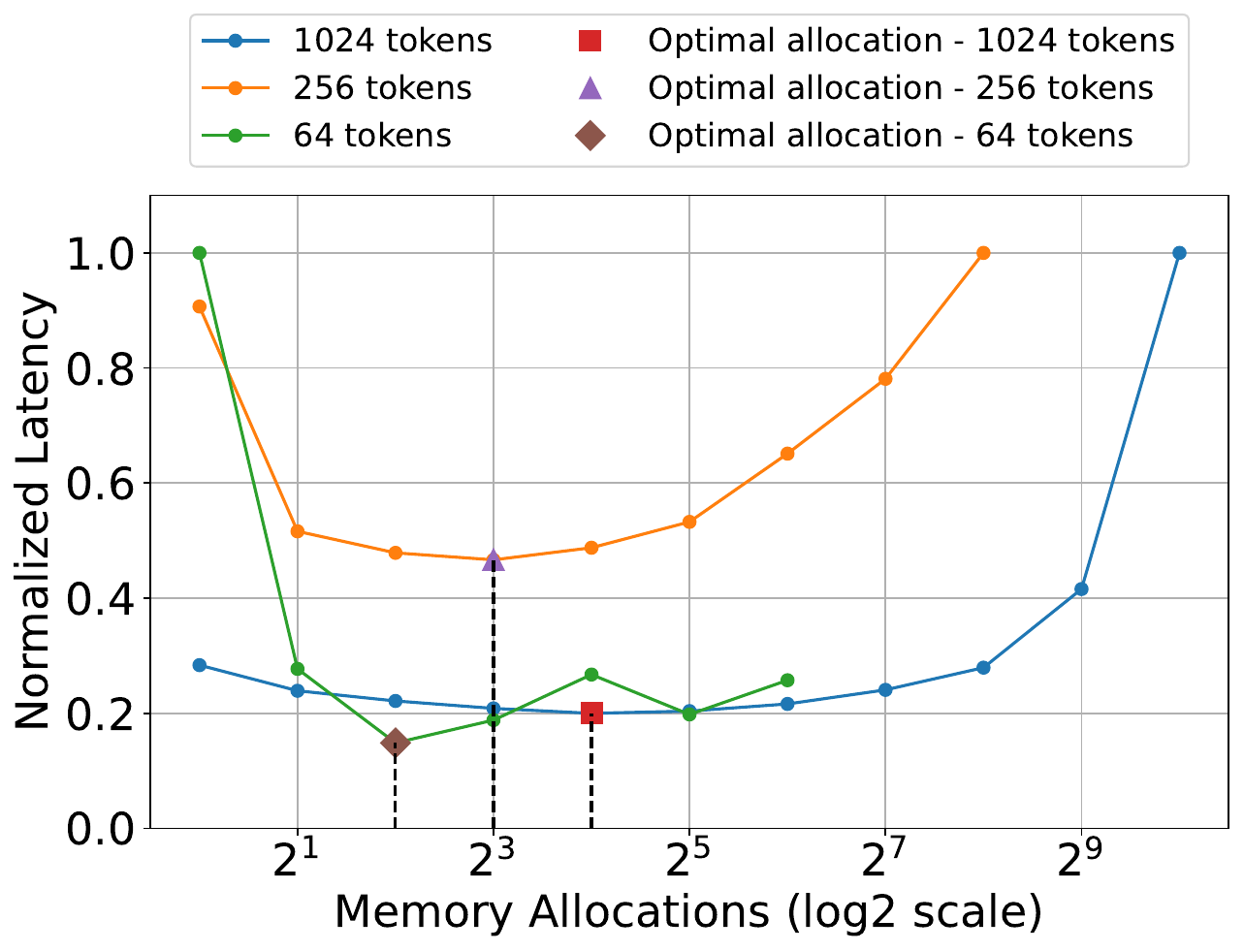}
	\captionof{figure}{Attention Block Latency on AMD Ryzen$^{TM}$ Desktop.}
         \label{sub:Ryzen}
    \end{minipage}
    \end{figure}
}

\subsubsection {Token Latency versus Context Length}

Figure ~\ref{E2E_CL} plots the latency of generating the $n$th token, 
for the baseline and the {\bmc} approach for the OPT-2.7B model with a batch size of 32.  
As context length increases, the latency of generating the next token increases significantly, roughly by 6$\times$ for $N$ going from $256$ to $2048$, with the iterative allocation scheme. 
Whereas for the {\bmc} scheme, the increase is relatively smaller (less than 2$\times$). The spikes in latency for the {\bmc} scheme occur at $(i{\cdot}r)$-th iteration, where memory allocation  and the associated page fault overheads are incurred.

\begin{figure}[htb]
\centering 
         \includegraphics[height=2.5cm]{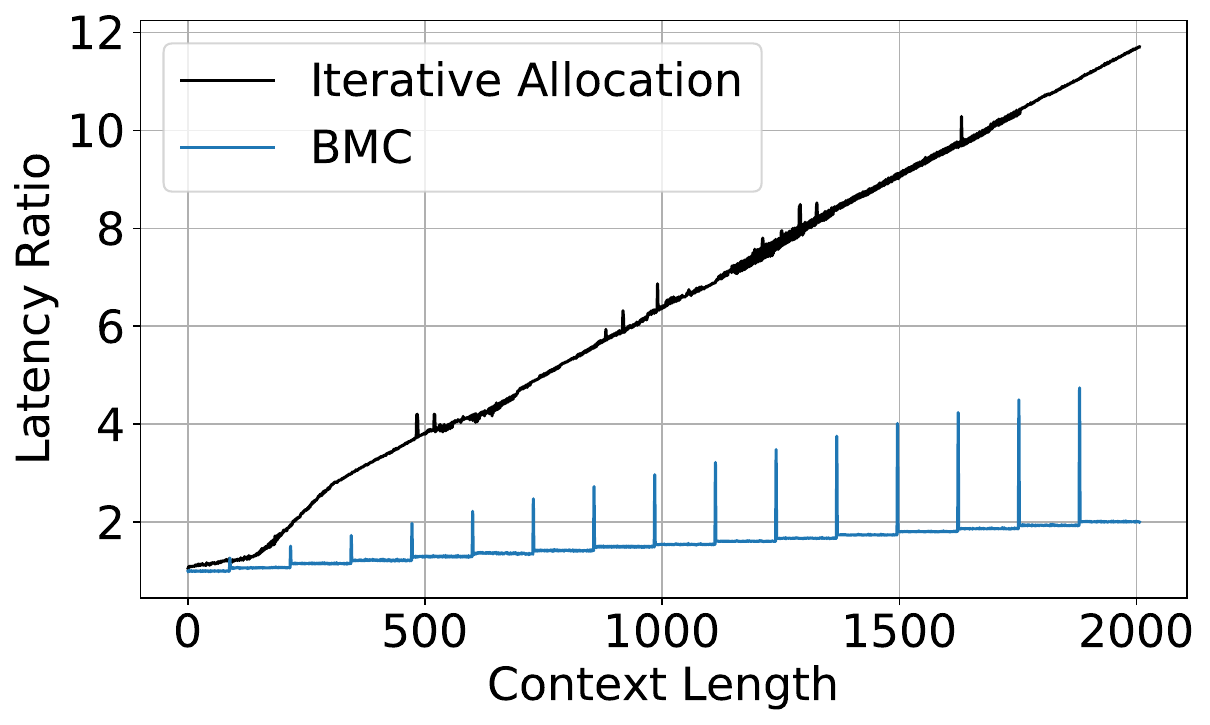}
	\caption{Latency of Generating $n$th Token} 
        \label{E2E_CL}       
    \end{figure}

\rnew{
\subsubsection{Alternative Approaches} 
\label{sec:strided_BMM}
\revise{Can KV cache copying overheads be reduced by upfront allocation and strided matrix multiplication?}{Are there other alternatives to eliminate/reduce  KV cache copying overheads?  
One possibility is to allocate KV cache upfront and perform matrix multiplication in a strided manner.} 
\revise{}{More specifically, the multihead attention computation,  as mentioned in Section~\ref{sec:MHA}, happens as a batch matrix multiplication of $Q=[B{\cdot}H,1,D]$ with the transpose of $K=[B{\cdot}H,n,D]$,  with the tensor dimension increasing with the context length $n$. 
With upfront allocation 
of size $[B{\cdot}H,N,D]$,  a GEMM call with appropriate stride is made to perform the batch matrix multiplication of $Q=[B{\cdot}H,1,D]$ with the transpose of $K=[B{\cdot}H,n,D]$. How does the strided version perform compared to one where the appropriate tensor size ($[B{\cdot}H,n,D]$) is allocated? }
For this experiment, we leverage oneDNN APIs~\cite{intel_onednn_2024}, and experiment for various tensor dimensions with $n=1536$ and $N=2048$. Our results, 
demonstrate that the strided version performs significantly worse, degradation ranging from \todo{3.01$\times$ to 14.4$\times$} \revise{(details not shown here due to space restriction)}{} over batch matrix multiplication on tensors of appropriate size.

}

\section{Conclusion}
In this paper, we propose a simple scheme called \emph{Balancing Memory and Compute} ({\bmc}), which trades off memory operations cost for some redundant computation to address inefficiencies in KV Cache operations.  
With speculative decoding, the redundant computation is repurposed to further improve the performance of {\bmc}. A simple analytical model identifies
best-performing design points that reduce the  computation time of attention block. \ignore{ The proposed approach is BLAS friendly.}
Empirical evaluation of the proposed method, \ignore {across a range of LLM models and inference configurations for a variety of CPU target platforms,} demonstrates \ignore { that BMC can integrate seamlessly with frameworks like HuggingFace and inference servers such as DeepSpeed, providing acceleration over the baseline. It also achieves} the proposed method is hardware agnostic. We also demonstrate  noticeable gains over  \ignore{inference servers like} vLLM and DeepSpeed. \ignore {\rnew{Additionally we demonstrate performance gains with Speculative Decoding and demonstrate the model is hardware agnostic.}}


\bibliographystyle{IEEEtranS}
\bibliography{sample-base}


\end{document}